\documentclass{ws-ijmpd}
\usepackage[super,compress]{cite}
\usepackage{color}
\usepackage{soul}

\usepackage{amsmath,amssymb}

\begin{document}

\markboth{A. Mart\'{i}n-Ruiz and L. F. Urrutia}
{Gravitational waves propagation in nondynamical Chern-Simons gravity}

\catchline{}{}{}{}{}

\title{Gravitational waves propagation in nondynamical Chern-Simons gravity}
\author{A. Mart\'{i}n-Ruiz${}^{*}$  and L. F. Urrutia${}^{\dagger}$}
\address{Instituto de Ciencias Nucleares, Universidad Nacional Aut\'{o}noma de M\'{e}xico, 04510 M\'{e}xico, Distrito Federal, M\'{e}xico \\  ${}^{*}$alberto.martin@nucleares.unam.mx \\ ${}^{\dagger}$urrutia@nucleares.unam.mx}

\maketitle

\begin{history}
\received{Day Month Year}
\revised{Day Month Year}
\end{history}

\begin{abstract}
 We investigate the propagation of gravitational waves in linearized Chern-Simons (CS) modified gravity by considering two nondynamical models for the coupling field $\theta$: (i) a domain wall and (ii) a surface layer of $\theta$, motivated by their relevance in condensed matter physics. We demonstrate that the metric and its first derivative become discontinuous for a domain wall of $\theta$, and we determine the boundary conditions by realizing that the additional contribution to the wave equation corresponds to one of the self-adjoint extensions of the D'Alembert operator. Nevertheless, such discontinuous metric  satisfies the area matching conditions introduced by Barrett. On the other hand, the propagation through a surface layer of $\theta$ behaves similarly to the propagation of electromagnetic waves in CS extended electrodynamics. In both cases we calculate the corresponding reflection and transmission amplitudes. As a consequence of the distributional character of the additional terms in the equations that describe wave propagation, the results obtained for the domain wall are not reproduced when the thickness of the surface layer goes to zero, as one could naively expect. 
\end{abstract}

\keywords{Chern-Simons gravity; gravitational waves; point interactions.}

\ccode{PACS numbers: 04.50.Kd, 04.30.-w, 04.60.Bc}


\section{Introduction}

One of the possible low-energy consequences of string theory is the addition of a Chern-Simons (CS) term coupled to an scalar field $\theta$ to the Einstein-Hilbert action of general relativity. Recently it has been suggested that this model provides the correct low-energy effective action to describe the thermal (gravitational) response of topological superconductors (TSCs) and superfluids, with the coupling field $\theta$ characterizing its topological nontriviality. CS gravity is an appealing topological extension of general relativity, where the Einstein-Hilbert action is supplemented with the parity-violating Pontryagin invariant coupled to an scalar field $\theta$, which can be considered either as a dynamical field or as an external prescribed quantity of spacetime \cite{Jackiw}. In much of the work on CS gravity, $\theta$ is assumed to be spatially homogeneous but time varying. This assumption can be motivated by arguments analogous to those that have been made suggesting that the quintessence field should be coupled to the electromagnetic CS term \cite{Smith, Lue, Yunes}.

Recently, CS theories have acquired great interest in condensed-matter (CM) physics, since they describe the topological response theories of topological insulators (TIs) and TSCs. TIs display nontrivial topological order and are characterized by a fully insulating bulk and gapless edge or surface states, which are protected by time-reversal symmetry \cite{Qi-Rev, Hasan-Rev}. TSCs are analog of TIs in superconductors, which have a full superconducting gap, and gapless edge states propagating on the boundary. In the case of the three-dimensional (3D) time-reversal invariant TIs, the topological response theory is described by the additional term in the electromagnetic action: $(\alpha / 32 \pi ^{2} ) \int \theta \epsilon _{\mu \nu \alpha \beta} F ^{\mu \nu} F ^{\alpha \beta} d ^{4} x$, in which $\theta$ is quantized to be $0$ (topologically trivial state) or $\pi$ mod $2 \pi$ (topologically nontrivial state). Such a topological term has the same form as the coupling of axions with gauge fields proposed in high-energy physics \cite{Peccei, Wilczek}. However, in a TI, $\theta$ is a constant determined by the bulk topology rather than a dynamical field. Many interesting properties for a domain wall of $\theta$ have been highlighted. For example, the image magnetic monopole effect \cite{Qi}, the topological Kerr and Faraday rotations \cite{Karch, Hehl, Zanelli, Huerta} arising from electromagnetic waves propagating through the $\theta$ boundary, and the quantized Hall effect \cite{Qi-Rev, Hasan-Rev}. A general technique to analyze the electromagnetic response of TIs has been elaborated in terms of Green's functions \cite{MCU1, MCU2, MCU3, MCU4}.

In the case of TSCs, since charge and spin are not conserved, the electromagnetic response is not adequate for defining the topological response theory. However the coupling to gravitational field, which describes the quantized thermal response, can be a good probe for topological nontriviality because energy is still conserved. It has been suggested that, in 3D time-reversal invariant TSCs and superfluids, one possible candidate for the topological response theory is the gravitational analog of the $\theta$ term for a gauge field \cite{TSC-Wang, TSC-Ryu, TSC-Qi, GEM-Furusaki, GEM-Nomura, GEM-Shiozaki}, i.e. a term proportional to the Pontryagin invariant of the Riemann curvature: 
\begin{align}
\mathcal{S} _{\scriptsize \mbox{CM}} = \frac{1}{1536 \pi ^{2}} \int \theta _{CM} \epsilon ^{\mu \nu \alpha \beta} R ^{\sigma} _{\phantom{\sigma}  \tau \alpha \beta} R ^{\tau} _{\phantom{\tau} \sigma \mu \nu} d ^{4} x , \label{ActTop}
\end{align}
with the topological order parameter being $\theta _{CM} = 0$ or $\pi$ mod $2 \pi$. It has been argued that the microscopic source of this gravitational response is the energy flow at the surfaces of a topological phase \cite{GEM-Furusaki, GEM-Nomura, GEM-Shiozaki, GEM-Sekine}. This is based upon the fact that temperature couples to the local energy density in the same way as an applied gravitational potential \cite{Luttinger}.

Motivated by  the steps followed in the study of the electromagnetic CS coupling, the main goal of this paper is to analyze the propagation of gravity waves in two models of  nondynamical CS gravity which have been shown to be relevant in CM physics. Since in this scheme $\theta$ characterizes the topological nontriviality of the media, we consider two nondynamical $\theta$-models which interpolate between two constant values of $\theta$: (i) a domain wall and (ii) a surface layer of $\theta$. In these scenarios, the gravitational field equations acquire additional contributions arising from the coupling between the Pontryagin density and the topological order parameter $\theta$. 

For a domain wall of $\theta$, these new distributional contributions have support only on the interface, while the usual Einstein field equations hold in the bulk. In this way, we have to deal with the problem of determining the right junction conditions to be imposed on the bulk metrics, such that their union form a valid solution to the field equations. This problem has been extensively discussed in the literature \cite{Israel, Taub, Raju, Geroch, Clarke, Hogan, Letelier, Podolsky, Mars} and arises in physical situations like the study of line sources, shock waves and thin shells of matter, for example. In this work the boundary conditions are not imposed by hand but they are dictated by the singular contributions to the field equations. Indeed, in the linear approximation, we find that the additional contribution to the gravitational  wave equation, induced by a delta distribution together with its derivative, corresponds to one of the self-adjoint extensions (SAE) of the D'Alembert operator. The general distributional analysis of this problem provides the junction conditions on the metric perturbation and its derivative at the domain wall \cite{Kurasov, Albeverio}. We find that even though the metric and its first derivative become discontinuous at the interface, the area matching condition introduced by Barrett \cite{Barrett} is satisfied. This amounts to replace the standard junction condition requiring the continuity of the metric at a given null hypersurface by the weaker condition that the area of any two-surface gives a unique result when measured from each side of the hypersurface. 

On the other hand, the case for a surface layer of $\theta$ is simpler since the additional contribution appearing in the field equation depends only on delta distributions supported at the interfaces. This case has a close resemblance with the propagation of electromagnetic waves in CS electrodynamics. We show that the results obtained for a surface layer do not reduce to the ones obtained with a domain wall model of $\theta$ . This is because, even when the surface layer become a domain wall when its thickness goes to zero, the limit in the field equations does not reproduces those of the domain wall.

This paper is organized as follows. We begin in Sec. \ref{CSGRAVITY} by reviewing the basics of nondynamical CS gravity following closely Ref.~\refcite{Jackiw}. We also discuss the linear theory and we demonstrate its consistency. In Sec. \ref{theta-models}, we introduce the nondynamical models for the coupling field $\theta$, we work with. Section \ref{GW} is devoted to the analysis of gravitational waves propagating across our $\theta$-models. We derive the boundary conditions by interpreting the field equations in a distributional sense. See also Sec. \ref{Discussion} and the \ref{App-SAE}. In Sec. \ref{phenomenological} we provide some phenomenological estimations of the parameters in our model. We include the comparison with two condensed matter representatives which are particular cases of our general discussion of wave propagation, once the scales are properly adjusted \cite{VolovikBook}. Finally, our summary and conclusions are given in Sec. \ref{Discussion}. Our conventions are those of Ref.~\refcite{Schutz}. We use the signature $ \left( - , + , + , + \right)$, the Riemann tensor is $R ^{\mu} _{\phantom{\mu} \nu \alpha \beta } = \partial _{\alpha} \Gamma ^{\mu} _{\nu \beta} + \Gamma ^{\mu} _{\sigma \alpha} \Gamma ^{\sigma} _{\nu \beta} - (\alpha \leftrightarrow \beta )$, the Ricci tensor is $R _{\mu \nu} = R ^{\alpha} _{\phantom{\alpha} \mu \alpha \nu}$ and $R = R ^{\mu} _{\phantom{\mu} \mu}$ is the Ricci scalar.

\section{Nondynamical CS Gravity}

\label{CSGRAVITY}

\subsection{Basics}

Let us consider the spacetime $\left( \mathcal{M} , g \right)$, with $\mathcal{M}$ a four-dimensional manifold and $g$ a metric on $\mathcal{M}$. The action of the nondynamical CS modified gravity reads
\begin{align}
\mathcal{S} = \mathcal{S} _{\scriptsize \mbox{EH}} + \mathcal{S} _{\scriptsize \mbox{CS}} + \mathcal{S} _{\scriptsize \mbox{matt}} , \label{Action}
\end{align}
where
\begin{align}
\mathcal{S} _{\scriptsize \mbox{EH}} = \frac{1}{2 \kappa } \int _{\mathcal{V}} d ^{4} x \sqrt{g} R \quad , \quad \mathcal{S} _{\scriptsize \mbox{CS}} = \frac{1}{2 \kappa} \int _{\mathcal{V}} d ^{4} x \frac{\theta }{4}\mathcal{P} .
\label{ACTIONS}
\end{align}
Here, $\kappa = 8 \pi G$, $\mathcal{V}$ is a four-volume in $\mathcal{M}$ with boundary $\partial \mathcal{V}$, $g$ is the determinant of the spacetime metric $g _{\mu \nu}$, $R$ is the Ricci scalar, $\theta$ is the CS scalar field, and $\mathcal{S} _{\scriptsize \mbox{matt}} (\Psi)$ is the matter action which does not depend on $\theta$. The field $\Psi$ collectively denotes the nongravitational matter fields. The quantity $\mathcal{P}$ in the CS action is the Pontryagin invariant defined as
\begin{equation}
\mathcal{P} = {}^{\ast} R _{\phantom{\sigma} \tau }^{\sigma \phantom{\tau} \mu \nu} R ^{\tau} _{\phantom{\tau} \sigma \mu \nu},  \label{Pontryagin}
\end{equation}
where ${}^{\ast } R  ^{\sigma \phantom{\tau} \mu \nu} _{\phantom{\sigma} \tau } = \frac{1}{2} \epsilon ^{\mu \nu \alpha \beta} R ^{\sigma} _{\phantom{\sigma}  \tau \alpha \beta}$ is the dual of the Riemann tensor and $\epsilon ^{\mu \nu \alpha \beta}$ is the Levi-Civit\`{a} symbol, with the convention $\epsilon ^{0123} = +1$.

The Pontryagin density can be expressed as a divergence $\nabla _{\mu} K ^{\mu} = \mathcal{P}$, where  
\begin{equation}
K ^{\mu} = \epsilon ^{\mu \nu \alpha \beta} \Gamma _{\nu \sigma }^{\lambda} \left( \partial _{\alpha} \Gamma _{\beta \lambda} ^{\sigma } +\frac{2}{3} \Gamma _{\alpha \xi} ^{\sigma} \Gamma _{\beta \lambda} ^{\xi} \right) , \label{CScurrent}
\end{equation}
is the  the CS topological current and $\Gamma $ is the Christoffel connection. Accordingly, if $\theta $ is globally constant at all spacetime points, the CS action is a topological term not contributing to the field equations since its variation is a boundary term that can be dropped under the usual boundary conditions.

The field equations of the nondynamical CS gravity are obtained by varying the action (\ref{Action}) with respect to the metric. One finds
\begin{equation}
G ^{\mu \nu} + C ^{\mu \nu} = \kappa T ^{\mu \nu} ,  \label{Field-Eqs}
\end{equation}
where $G ^{\mu \nu} = R ^{\mu \nu} - \frac{1}{2} g ^{\mu \nu} R $ is the covariantly conserved Einstein tensor and $T ^{\mu \nu}$ is the matter stress-energy tensor. The expression for the symmetric traceless second rank tensor $C ^{\mu \nu}$, a four-dimensional (4D) generalization of the 3D Cotton-York tensor, is
\begin{equation}
C ^{\mu \nu} = - \frac{1}{2 \sqrt{g}} \left[ \upsilon _{\lambda} \epsilon
^{\lambda \mu \alpha \beta} \nabla _{\alpha} R ^{\nu} _{\;\ \beta} +
\upsilon _{\lambda \sigma} {}^{\ast} R ^{\sigma \mu \lambda \nu} + \left(
\mu \leftrightarrow \nu \right) \right],  \label{Cotton}
\end{equation}
where $\upsilon _{\lambda} = \nabla _{\lambda} \theta$ is called the embedding coordinate and $\upsilon _{\lambda \sigma} = \nabla
_{\sigma} \upsilon _{\lambda}$ is its covariant derivative.

By construction, the Einstein tensor $G _{\mu \nu}$ is divergenceless. If $\theta$ is treated as an external prescribed quantity, then general covariance, which requires $\nabla _{\mu} T ^{\mu \nu} =0$, leads to the constraint $\nabla _{\mu} C ^{\mu \nu} = 0$. However, $C ^{\mu \nu}$ is not covariantly conserved. Rather, we have
\begin{equation}
\nabla _{\mu} C ^{\mu \nu} = \frac{\partial ^{\nu} \theta }{2 \sqrt{g}}
\mathcal{P} ,  \label{ConsistencyCond}
\end{equation}
and thus the consistency requirement on Eq. (\ref{Field-Eqs}) demands $\mathcal{P} = 0$. Alternatively, if we treat $\theta$ as a dynamical field, then the variation of the action with respect to $\theta$ will lead to the same constraint on $\mathcal{P}$ \cite{Jackiw}. In this paper, we are interested in the propagation of gravitational waves, where $T _{\mu \nu} = 0$, and the constraint $\nabla _{\mu} C ^{\mu \nu} = 0$ is satisfied regardless we view $\theta$ as a dynamical field or a fixed, externally specified quantity, because the Pontryagin density is identically zero for GW spacetimes \cite{Jackiw}.

\subsection{Linear theory} \label{LinearCS}

We work in the weak field approximation of the nondynamical CS gravity. In the linear approximation, $g _{\mu \nu}= \eta _{\mu \nu} + \varepsilon h _{\mu \nu}$, the source free field equation to first order in $\varepsilon \ll 1$ is $G ^{\mu \nu} _{\scriptsize \mbox{linear}} + C ^{\mu \nu} _{\scriptsize \mbox{linear}} = 0$, where
\begin{align}
G ^{\mu \nu} _{\scriptsize \mbox{linear}} &= \frac{1}{2} \Big[ - \Box h ^{\mu \nu} + \partial _{\alpha} \partial ^{\mu} h ^{\nu \alpha} + \partial _{\alpha} \partial ^{\nu} h ^{\mu \alpha} - \partial ^{\mu} \partial ^{\nu} h - \eta ^{\mu \nu} \left( \partial ^{\alpha} \partial ^{\beta} h _{\alpha \beta} - \Box h \right) \Big] , \label{Einstein-Linear} \\ C ^{\mu \nu} _{\scriptsize \mbox{linear}} &= \frac{1}{4} \Big[ \upsilon _{\lambda} \epsilon ^{\lambda \mu \alpha \beta} \partial _{\alpha} \left( \Box h ^{\nu} _{\phantom{\nu} \beta} - \partial ^{\kappa} \partial ^{\nu} h _{\beta \kappa} \right) + \upsilon _{\sigma \lambda} \epsilon ^{\sigma \mu \alpha \beta} \partial _{\alpha} \left( \partial ^{\lambda} h ^{\nu} _{\phantom{\nu} \beta} - \partial ^{\nu} h ^{\lambda} _{\phantom{\lambda} \beta} \right)  + ( \mu \leftrightarrow \nu ) \Big] , \label{Linear-Cotton-York}
\end{align}
are the linearized Einstein and Cotton-York tensors, respectively. Here, $\Box = \partial _{\mu} \partial ^{\mu}$ is the flat-space d'Alembertian, $h = h ^{\mu} _{\phantom{\mu} \mu}$ and indices are moved with $\eta ^{\mu \nu}$. The consistency condition requiring the vanishing divergence of the field equation can be directly verified. On the one hand, the Einstein tensor is naturally divergenceless, i.e. $\partial _{\mu} G ^{\mu \nu} _{\scriptsize \mbox{linear}} = 0$. On the other hand, using the antisymmetry of the Levi-Civit\`{a} symbol, the divergence of the Cotton-York tensor becomes
\begin{align}
\partial _{\mu} C ^{\mu \nu} _{\scriptsize \mbox{linear}} = \frac{1}{4} \Big[ \upsilon _{\lambda \mu} \epsilon ^{\lambda \nu \alpha \beta} \partial _{\alpha} \left( \Box h ^{\mu} _{\phantom{\mu} \beta} - \partial ^{\kappa} \partial ^{\mu} h _{\beta \kappa} \right)  + \upsilon _{\sigma \lambda} \epsilon ^{\sigma \nu \alpha \beta} \partial _{\alpha} \partial _{\mu} \left( \partial ^{\lambda} h ^{\mu} _{\phantom{\mu} \beta} - \partial ^{\mu} h ^{\lambda} _{\phantom{\lambda} \beta} \right) \notag \\ + \upsilon _{\sigma \lambda \mu} \epsilon ^{\sigma \nu \alpha \beta} \partial _{\alpha} \left( \partial ^{\lambda} h ^{\mu} _{\phantom{\mu} \beta} - \partial ^{\mu} h ^{\lambda} _{\phantom{\lambda} \beta} \right) \Big] , \label{Linear-Cotton-YorkDMU}
\end{align}
with $\upsilon _{\sigma \lambda \mu} = \partial _{\mu} \upsilon _{\sigma \lambda} = \partial _{\sigma} \partial _{\lambda} \partial _{\mu} \theta$. The sum of the first two terms in the right-hand side  is zero since $\Box = \partial _{\mu} \partial ^{\mu}$, and the last term identically vanishes by symmetry considerations.

Next, we simplify the field equations by using the gauge freedom of the linearized approximation. We work with the trace-reversed metric perturbation, $\overline{h} _{\mu \nu} = h _{\mu \nu} - \frac{1}{2} \eta _{\mu \nu} h$, and impose the Lorentz gauge condition, $\partial ^{\mu} \overline{h} _{\mu \nu} = 0$ to obtain the linearized tensors
\begin{align}
G ^{\mu \nu} _{\scriptsize \mbox{linear}} &= - \frac{1}{2} \Box \overline{h} ^{\mu \nu}  , \label{Linear-Einstein} \\ C ^{\mu \nu} _{\scriptsize \mbox{linear}} &= \frac{1}{4} \Big[ \upsilon _{\lambda} \epsilon ^{\lambda \mu \alpha \beta} \partial _{\alpha} \Box \overline{h} ^{\nu} _{\phantom{\nu} \beta} + \upsilon _{\sigma \lambda} \epsilon ^{\sigma \mu \alpha \beta} \partial _{\alpha} \left( \partial ^{\lambda} \overline{h} ^{\nu} _{\phantom{\nu} \beta} - \partial ^{\nu} \overline{h} ^{\lambda} _{\phantom{\lambda} \beta} \right) + ( \mu \leftrightarrow \nu ) \Big] . \label{Linear-Cotton}
\end{align}
This choice reduces the number of degrees of freedom from 10 to 6. A further gauge transformation can be made. In the next section, we will show that the TT gauge can be imposed to analyze the propagation of gravitational waves in our models, reducing the number of degrees of freedom from 6 to 2, as it should be.

\section{Nondynamical $\theta$-models} \label{theta-models}

One of the main ingredients of nondynamical CS gravity is the CS coupling scalar $\theta = \theta (x ^{\mu})$, which is a prescribed function of spacetime. In the nondynamical framework, the functional form of the CS scalar field is usually taken to be dependent only of time, $\theta (t) = \tau t$, where $\tau$ is a constant. The assumption being that $\theta$ is a quintessence field or some other field that somehow echoes the arrow of time associated with the cosmic expansion. 

In this paper, we are concerned with nondynamical $\theta$-models inspired by CM systems. We begin by recalling the emergence of the CS term in the theory of TIs and TSCs. The magnetoelectric polarization $\theta$ characterizing the momentum-space topology of band insulators gives rise to CS electrodynamics, in which  $\theta$ couples with the electromagnetic field via $ \frac{\alpha}{16 \pi ^{2}} \theta {\;}^{\ast} F ^{\mu \nu} F _{\mu \nu}$. A domain wall of $\theta$ leads to the quantized Hall effect and the image monopole effect. In the case of TSCs, the scalar field $\theta$, associated with the topological nontriviality of these states, gives rise to CS gravity, in which $\theta$ couples to the gravitational field via $\frac{1}{768 \pi ^{2}} \theta \mathcal{P}$, where $\mathcal{P}$ is the gravitational Pontryagin density (\ref{Pontryagin}). The microscopic source of this gravitational response is the energy flow at surfaces of a topological phase. According to the Luttinger derivation of the thermal transport coefficients, in a near-equilibrium system, the effect of a thermal gradient is equivalent to that obtained from a gravitational potential \cite{Luttinger}. Consequently, since the surface Majorana mode that exists in this phase does not carry charge, but it does carry energy, it leads to a thermal Hall effect for a domain wall of $\theta$. 

In analogy with the electromagnetic case, the propagation of gravitational waves in the models described above is a topic which deserves to be studied on its own. In this work, we shall consider that the CS scalar field depends only of one spatial coordinate, which we choose to be $z$, i.e. $\theta = \theta (z)$. Therefore, the embedding coordinate and its derivative become $\upsilon _{\mu} = \theta ^{\prime} n _{\mu}$ and $\upsilon _{\mu \nu} = \theta ^{\prime \prime} n _{\mu} n _{\nu}$, respectively. Here, $n _{\mu}=(0,0,0,1)$ is the unit vector pointing in the direction of the coordinate $z$ and the primes denote derivatives with respect to $z$.

The linearized field equations can be further simplified imposing an additional gauge. Hereafter, we work in the transverse-traceless gauge, defined by the (previously imposed) Lorentz gauge together with the conditions $\overline{h} ^{\mu} _{\phantom{\mu} \mu} = 0$ and $U ^{\mu} \overline{h} _{\mu \nu} = 0$, with  $U^{\mu }$ being any constant timelike four vector. With the appropriate choice of the four vector $U ^{\mu}$ we can make $n ^{\mu} \overline{h}  _{\mu \nu} = 0$. For example, for monochromatic plane waves with four momentum $k ^{\mu} = \omega \hat{k} ^{\mu} = \omega (1, \hat{\textbf{k}})$, the choice $U ^{\mu} = \hat{k} ^{\mu} - (\hat{k} \cdot n) n ^{\mu}$ implies $n ^{\mu} \overline{h}  _{\mu \nu} = 0$, when $k\cdot n \neq 0$. Then, the field equations reduce to
\begin{align}
\Box \overline{h} ^{\mu \nu} = \frac{1}{2}  \epsilon ^{3 \mu \alpha \beta} \Big( \theta ^{\prime} \Box + \theta ^{\prime \prime} \partial _{z} \Big) \partial _{\alpha} \overline{h} ^{\nu} _{\phantom{\nu} \beta} + ( \mu \leftrightarrow \nu ) . \label{Field-Eq-Fin}
\end{align}
Since the contributions of the $\theta$-term to the field equations can be seen as a source of energy-momentum, which nevertheless arises from the fields themselves, it is important to verify that there are no obstructions to attain the TT gauge. Let us recall that in the standard case of gravitational wave propagation in the presence of matter characterized by an energy-momentum tensor $T^{\mu\nu}$, it is always possible to implement the Lorentz gauge (since $\partial_\mu T^{\mu\nu}=0$), but not the subsequent TT gauge conditions. This is because, in general, we have $T^{\mu}_{\phantom{\mu}\mu} \neq 0$ and $n_\mu T^{\mu\nu} \neq 0$, which will make the equation of motion inconsistent after imposing the TT gauge \cite{Padmanabhan}. In our case, this does not happen because the effective source of energy-momentum is given by $\overline{h} ^{\mu \nu}$. In fact, Eq. (\ref{Field-Eq-Fin}) is consistent with the TT gauge conditions $ h^{\mu}_{\phantom{\mu} \mu}=0, \,  {\overline{h}}^{3\nu}=0 , \, \partial_\mu \overline{h}^{\mu \nu}=0$. One can verify that: (i) The traceless condition arises from the symmetry properties in the indices $\mu=\nu, \beta$. (ii) Equation (\ref{Field-Eq-Fin}) vanishes when projected along $n _{\mu}$. The first term in the right-hand side is zero due to $\epsilon^{3 3 \alpha \beta}=0$ and the second term contains $\overline{h}^{3}_{\phantom{3}\beta}=0$. (iii) When applying $\partial_\mu$ upon Eq. (\ref{Field-Eq-Fin}), the Lorentz condition is satisfied because the first term in the right-hand side is zero, since $\epsilon^{3\mu\alpha\beta}$ restricts both $\mu$ and $\alpha$ to live in the subspace orthogonal to $n_\mu$. In this way, $\partial_\mu$ commutes with the operator in square brackets (which depends only on $z$)  yielding  $\partial_\mu\partial_\alpha \overline{h}^{\nu}_{\phantom{\nu} \beta}$, which is zero due to  the antisymmetry of the Levi-Civit\`{a} symbol. For analogous reason, the second term  yield the contribution $\partial_\mu \overline{h}^{\mu}_{\phantom{\mu} \beta}=0$.

Next we define the particular nondynamical $\theta$-models we deal with: (i) a domain wall and (ii) a surface layer of $\theta$. 

\subsection{Domain wall model}

\begin{figure}[ht]
\centerline{ \psfig{file=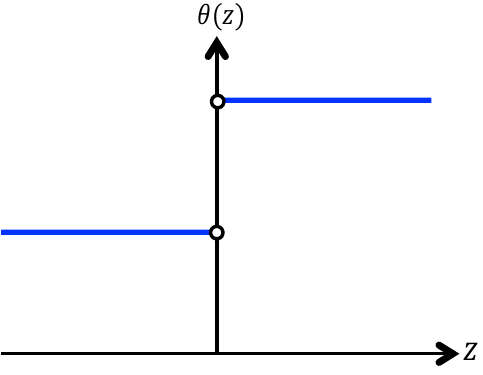,width=5cm}}
\vspace*{8pt}
\caption{Domain wall of $\theta$.  \label{Mod1}}
\end{figure}

We first consider a domain wall between two different constant values of $\theta$, lying along the $xy$-plane at $z=0$, as shown in Fig. \ref{Mod1}. More specifically, we assume that the region $z < 0$ is characterized by a constant $\theta=\theta _{1}$, while the region $z > 0$ has a different
constant value $\theta=\theta _{2}$. 

The invariance of the CS term in the action under shifts of $\theta$  ￼￼￼by any constant, $\theta = \theta ^{\prime} + C$, can be used to set $\theta _{1}$ to zero and $\theta _{2}$ to ${\tilde \theta} \equiv \theta _{2} - \theta _{1}$. In this way, the derivatives of the CS scalar field appearing in Eq. (\ref{Field-Eq-Fin}) become $\theta ^{\prime} = {\tilde \theta} \delta (z)$ and $\theta ^{\prime \prime} = {\tilde \theta} \delta ^{\prime} (z)$, and the field equation (\ref{Field-Eq-Fin}) reads
\begin{align}
\Box \overline{h} ^{\mu \nu} = \frac{{\tilde \theta}}{2} \epsilon ^{3 \mu \alpha \beta} \Big[ \delta (z) \Box + \delta ^{\prime} (z) \partial _{z} \Big] \partial _{\alpha} \overline{h} ^{\nu} _{\phantom{\nu} \beta} + ( \mu \leftrightarrow \nu ) . \label{Field-Eq-Mod1}
\end{align}
The presence in the above equation of the Dirac delta distribution and its derivative implies that the right-hand side is supported only at the interface $z=0$ and vanishes in the bulk. Therefore, the metric perturbation satisfies the usual wave equation in the bulk regions, $z<0$ and $z>0$, and the domain wall acts as a localized source depending on the gravitational field itself. The $\delta ^{\prime}$ distribution appearing in Eq. (\ref{Field-Eq-Mod1}) is a consequence of the third derivatives of the metric tensor arising from the CS contribution, and it implies that both, (or at lest some components of) the metric perturbation and its first derivative become discontinuous at the interface. Here, the field equations dictate the junction conditions for the metric at the interface.

\subsection{Surface layer model}

\begin{figure}[!ht]
\centerline{ \psfig{file=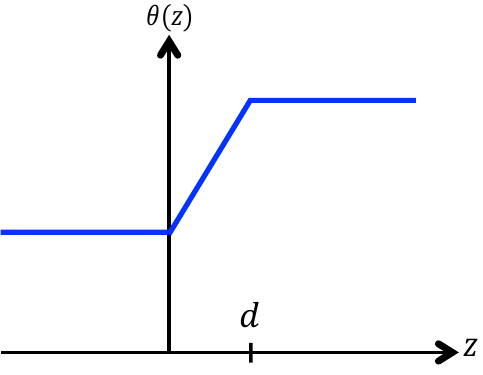,width=5cm}}
\vspace*{8pt}
\caption{Surface layer of $\theta$.}
\label{Mod2}
\end{figure}

The second model we shall consider consists of a surface layer where $\theta$ changes linearly across a thickness $d$. More precisely, we assume that $\theta$ is constant in the region $z < 0$, then changes linearly in $z$ within the surface layer ($0 < z < d$) and finally becomes constant  again outside the layer ($z>d$), as shown in Fig. \ref{Mod2}.

Again, we can use the symmetry of the CS term under shifts of $\theta(z)$ by any constant to set $\theta_1 = 0$ for $z<0$ and ${\tilde \theta} \equiv \theta_2-\theta_1 $ for $z>d$. In this model, the derivatives of the CS scalar field become $\theta ^{\prime} = ( {\tilde \theta} / d ) \Pi (z)$ and $\theta ^{\prime \prime} = ({\tilde \theta} / d ) \left[ \delta (z) - \delta (z - d) \right] $, where $\Pi (z) = H(z) H (z-d)$ is the box function within the layer. The field equations now take the form
\begin{align}
\Box \overline{h} ^{\mu \nu} &= \frac{{\tilde \theta}}{2 d} \epsilon ^{3 \mu \alpha \beta} \Big\{ \Pi (z) \Box + \left[ \delta (z) - \delta (z - d) \right] \partial _{z} \Big\} \partial _{\alpha} \overline{h} ^{\nu} _{\phantom{\nu} \beta}  + ( \mu \leftrightarrow \nu ) . \label{Field-Eq-Mod2}
\end{align}
We observe that the first term in the right-hand side is nonzero within the layer, but vanishes outside it. The second term has support only  at the surfaces of the layer, that is to say at $z=0$ and $z=d$. Therefore, the above equation implies that the metric perturbation satisfies the usual wave equation $\Box \overline{h} ^{\mu \nu} =0$ outside the layer, and $\Box \overline{h} ^{\mu \nu} = ({\tilde \theta} / 2 d ) \epsilon ^{3 \mu \alpha \beta} \Pi (z) \partial _{\alpha} \Box \overline{h} ^{\nu} _{\phantom{\nu} \beta}$ in the interior region. In other words, the solution inside the layer can also be taken as satisfying the usual wave equation. 

\section{Propagation of Gravitational Waves}

\label{GW}

In this section, we examine the effects of our nondynamical $\theta$-models in the propagation of gravitational waves. To begin with let us consider an incident monochromatic plane wave impinging from the left towards the $\theta$-interface at $z=0$, at an angle $\alpha$ with respect to the normal $n ^{\mu} = (0,0,0,1)$, with wave four-vector $k ^{\mu} = \omega \hat{k} ^{\mu} = \omega \left( 1 , \sin \alpha , 0 , \cos \alpha \right)$. The constant timelike vector which fixes the TT gauge is $U ^{\mu} = ( 1, \sin \alpha , 0 , 0)$, with $0 \leq \alpha \leq \pi /2$. In this gauge, the metric perturbation $h _{\mu \nu}$ has the form
\begin{equation}
\overline{h} _{\mu \nu} \left( t , x , z \right) = \tilde{e} _{\mu \nu} ^{(+)} h _{+}\left( t , x , z \right) + \tilde{e} _{\mu \nu} ^{(\times)} h _{\times}\left(  t , x ,z \right) .  \label{GW-Gen-Ansatz}
\end{equation}
Here $h _{+}$ and $h _{\times}$ are the amplitudes of the two independent components with linear polarizations, and
\begin{align}
\tilde{e} _{\mu \nu} ^{(+)} &= \left(
\begin{array}{cccc}
\sin ^{2} \alpha & - \sin \alpha & 0 & 0 \\
- \sin \alpha & 1 & 0 & 0 \\
0 & 0 & - \cos ^{2} \alpha & 0 \\
0 & 0 & 0 & 0
\end{array} \right) \quad , \quad  \tilde{e}_{\mu \nu} ^{(\times)} &= \left(
\begin{array}{cccc}
0 & 0 & -\sin \alpha & 0 \\
0 & 0 & 1 & 0 \\
-\sin \alpha & 1 & 0 & 0 \\
0 & 0 & 0 & 0
\end{array} \right) ,  \label{PolarizationTensors}
\end{align}
are the corresponding polarization tensors. The modified field equations (\ref{Field-Eq-Mod1}) and (\ref{Field-Eq-Mod2}) can be written in the generic form
\begin{align}
\Box \overline{h} _{\mu \nu} = - i {\tilde \theta} \hat{\mathcal{O}} \left[ \tilde{e} _{\mu \nu} ^{(+)} h _{\times} - \cos ^{2} \alpha \; \tilde{e} _{\mu \nu} ^{(\times)} h _{+} \right] ,  \label{Field-Eq-Gen}
\end{align}
where the differential operator $\hat{\mathcal{O}}$ depends on the model under consideration. For the domain wall model, it is
\begin{align}
\hat{\mathcal{O}} _{\rm dw} = i \left[ \delta (z) \Box + \delta ^{\prime } (z) \partial _{z}\right] \partial _{t} , \label{O-dw}
\end{align}
while for the surface layer model it becomes
\begin{align}
\hat{\mathcal{O}} _{\rm sl} = \frac{i}{d} \Big\{ \Pi (z) \Box + \left[ \delta (z) - \delta (z - d) \right] \partial _{z} \Big\} \partial _{t} . \label{O-sl}
\end{align}
Equation (\ref{Field-Eq-Gen}) couples the polarization modes $+$ and $\times$, which are detached by introducing the right- and left-handed circularly polarized modes
\begin{align}
h _{R/L} = \cos \alpha \; h _{+} \pm i \; h _{\times} ,  \label{CircularPol}
\end{align}
that satisfy
\begin{align}
\Box h _{R/L} = \mp {\tilde \theta} \cos \alpha \; \hat{\mathcal{O}} h _{R/L} .  \label{Field-Eq-CircPol}
\end{align}
The plus and minus signs correspond to the right- and left-handed modes, respectively. Now we have to solve Eq. (\ref{Field-Eq-CircPol}) for the different models.

As discussed before, the differential operator in the right hand side of Eq. (\ref{Field-Eq-CircPol}) is singular, in the sense that it is supported at $z=0$ for the domain wall model, and at the surfaces (at $z=0$ and $z=d$) of the surface layer model, but vanishes in the bulk. This situation resembles the case of point interactions in nonrelativistic quantum mechanics, which are usually described by the so-called Fermi pseudopotentials. Point interactions are modeled by an idealized localized singular interaction with zero range occurring at one point on the space. One of these point interactions is the familiar $\delta \left( z \right) $ pseudopotential, which is well defined and has well-known solutions. This kind of interaction can be described by a free system on the line without the singular point, i.e. in the region $\mathbb{R} \setminus \left\{0\right\} $, in which case the interaction is encoded in the boundary conditions rather than in a formal Hamiltonian operator. For the special case of the $\delta$-interaction, the Schr\"{o}dinger equation can be integrated twice for obtaining the boundary condition at $z=0$, which demand: (1) continuity of the wave function and (2) discontinuity of its first derivative. However, attempts to consider more general interactions have been known to be plagued with difficulties associated with the definition of the interaction term. For example, there is some controversy on the meaning of the $\delta ^{\prime} \left( z \right)$-potential, as different regularization produce different reflection and transmission coefficients \cite{Griffiths, Coutinho, Seba, Christiansen, Toyama}. The origin of such difficulties lies in the fact that point interactions are represented by pseudopotentials which are not ordinary functions but distributions. 

A mathematically rigorous approach to deal with generalized point interactions in nonrelativistic quantum mechanics is due to Kurasov and coworkers in Refs.~\refcite{Kurasov}-\refcite{Albeverio}. In this approach, both the interaction and the discontinuous wave function are considered as distributions, precluding the naive definition of the interaction term as the usual product between the wave function and the potential energy, since such a product is generally ill-defined in distribution theory. Also, Kurasov deals with the problem of point interactions from the perspective that they are defined by SAE of the operator $-d^{2}/dz^{2}$, which is a well studied problem in the literature. This point of view has the advantage that the definition for the singular potentials depends only on the matching conditions at the singularity and not on the choice of a particular model of the delta function and its derivatives.

In the problem at hand, we have to deal with a second-order differential
equation for the metric, subject to the singular differential operator $\hat{\mathcal{O}}$ defined in Eqs. (\ref{O-dw}) and (\ref{O-sl}). The goal of the rest of this section is to analyze the one-dimensional (1D) differential equation (\ref{Field-Eq-CircPol}) following Kurasov's prescription, which is rephrased in terms of pseudopotentials and discontinuous functions in the \ref{App-SAE}. In this way, following Refs.~\refcite{Kurasov}-\refcite{Albeverio}, we are able to determine the boundary conditions for Eq. (\ref{Field-Eq-CircPol}) by standard manipulations of second-order linear differential equations.

To begin with we recall the expressions (\ref{deltaf22}), arising from the approach of Ref.~\refcite{Kurasov}, for the product of a function $\psi (x)$ which is discontinuous at $x=0$ times $\delta (x)$ and $\delta ^{\prime }(x)$:
\begin{align}
\label{DiscFuncDef} 
\begin{split}
\psi \left( x \right) \delta \left( x \right) &= \overline{\psi \left( 0 \right)} \delta \left( x \right) ,  \\ \psi \left( x \right) \delta ^{\prime} \left( x \right) &= \overline{\psi \left( 0 \right)} \delta ^{\prime} \left( x \right) - \overline{\psi ^{\prime} \left( 0 \right)} \delta \left( x \right) ,  \end{split}
\end{align}
with
\begin{equation}
\overline{\chi \left( 0 \right)} = \frac{1}{2} \left[ \chi \left( 0 ^{+}\right) + \chi \left( 0 ^{-} \right) \right] .   \label{AverageOrigin}
\end{equation}
Here $\chi$ denotes either $\psi $ or $\psi ^{\prime}$ and $\chi \left( 0 ^{+} \right)$ and $\chi \left( 0 ^{-} \right)$ are the limits of $\chi \left( x \right)$ when $x$ approaches $0$ from the positive and negative sides, respectively. In the next sections, we will tackle each of our models independently. First, we start with the modified field equations from the distributional point of view. Next, we establish the corresponding boundary conditions for the metric perturbation and its derivative, and finally we compute the reflection and transmission coefficients by using an appropriate ansatz for the solutions.

\subsection{Gravitational waves propagating through a domain wall} 
\label{Prop-DW}

We begin with the distributional differential equation (\ref{Field-Eq-CircPol}) describing the propagation through a domain wall. This is obtained by introducing the relations (\ref{DiscFuncDef}) into the modified wave equation (\ref{Field-Eq-CircPol}) with the point interaction (\ref{O-dw}). We obtain
\begin{align}
\Box h _{R/L}\left( z \right) &= \mp i {\tilde \theta} \cos \alpha \Big[ \delta \left( z \right) \overline{\partial _{t} (\Box - \partial _{z} ^{2}) h _{R/L} \left( 0 \right)} + \delta ^{\prime} \left( z \right) \overline{\partial _{z} \partial _{t} h _{R/L} \left( 0 \right)} \Big] , \label{GWs-dw}
\end{align}
where the overline denotes average at the domain wall, i.e. at $z=0$. Now we are in position to obtain the boundary conditions.

\subsubsection{Boundary conditions}

The first boundary condition for the metric perturbation at the domain wall can be obtained by integrating (\ref{GWs-dw}) over the interval $\left[ - \varepsilon , + \varepsilon \right]$ and taking the limit $\varepsilon \rightarrow 0^{+}$. The result is
\begin{equation}
h _{R/L} ^{\prime} \left( 0 ^{+} \right) - h _{R/L} ^{\prime} \left( 0 ^{-}\right) = \mp i {\tilde \theta} \cos \alpha \; \overline{\partial _{t} (\Box - \partial _{z} ^{2}) h _{R/L} \left( 0 \right)} ,
\label{BC1-dw}
\end{equation}
where the relations $\int _{-\varepsilon} ^{+ \varepsilon} \delta \left(
x \right) dx = 1$ and $\int _{- \varepsilon} ^{+ \varepsilon} \delta ^{\prime} \left( x \right) dx = 0$ have been used. Meanwhile, integrating (\ref{GWs-dw}) from $- L$ (with $L$ positive) to $z$ yields
\begin{align}
& h _{R/L} ^{\prime} \left( z \right) - h _{R/L} ^{\prime} \left( - L \right) - \partial _{t} ^{2} \int _{- L} ^{z} h _{R/L} \left( z ^{\prime} \right) dz ^{\prime} = \mp i {\tilde \theta} \cos \alpha \times \notag \\ & \hspace{3.5cm} \left[ H \left( z \right) \overline{\partial _{t} (\Box - \partial _{z} ^{2}) h _{R/L} \left( 0 \right)} + \delta \left( z \right) \overline{\partial _{z} \partial _{t} h _{R/L} \left( 0 \right)} \right] ,  \label{BC2.1-dw}
\end{align}
where $H \left( z \right) $ is the Heaviside function. Here, we have used the relations $\int _{- L} ^{z} \delta \left( y \right) dy = H \left( z \right) $ and $\int _{- L} ^{z} \delta ^{\prime} \left( y \right) dy =\delta \left( z \right)$. Integrating further (\ref{BC2.1-dw}) with respect to $z$ from $- \varepsilon $ to $+ \varepsilon $, and taking the limit $\varepsilon \rightarrow 0 ^{+}$, we find that
\begin{equation}
h _{R/L}\left( 0 ^{+} \right) - h _{R/L} \left( 0 ^{-} \right) = \mp i {\tilde \theta} \cos \alpha \; \overline{\partial _{z} \partial _{t} h _{R/L} \left( 0 \right)} ,
\label{BC2-dw}
\end{equation}
which is the second boundary condition for the metric perturbation at the domain wall.

\subsubsection{The transmission and reflection coefficients}

It is well known that the linearized Einstein field equations in vacuum ($T ^{\mu \nu} = 0$) far outside the source of the field has a (complex) solution of the form
\begin{equation}
h ^{\mu \nu} = A ^{\mu \nu} e^{i k _{\lambda} x ^{\lambda}},  \label{PlaneWaves}
\end{equation}
describing plane linearized gravitational waves, where $A ^{\mu \nu}$ are the (complex) constant amplitudes of the wave and $k ^{\mu}$ is the null wave four-vector given at the beginning of Sec. \ref{GW}. Here, we shall use this simple solution to illustrate the effect of a domain wall of $\theta$ in the propagation of linearized gravitational waves. Assuming that the source lies in the region $z<0$, the metric perturbation impinging on the surface $z=0$ from the left can thus be written as
\begin{align}
h _{R/L} (x,z,t) = \left\lbrace \begin{array}{c} e ^{i k _{\mu} x ^{\mu}} + \mathcal{R} _{R/L} e ^{i \tilde{k} _{\mu} x ^{\mu}} \\[7pt] \mathcal{T} _{R/L} e ^{i k _{\mu} x ^{\mu}} \end{array} \right. \begin{array}{c} ; \quad z < 0 \\[7pt] ; \quad z > 0 \end{array}  \label{metricGW}
\end{align}
The four-vector $\tilde{k} ^{\mu} = \omega (1 , \sin \alpha , 0 , - \cos \alpha)$ is the wave vector of the reflected wave, and $\mathcal{R}_{R/L}$ and $\mathcal{T}_{R/L}$ are the corresponding reflected and transmitted amplitudes. By using the simple results $\partial _{t}h_{R/L}=- i \omega h _{R/L}$ and $\partial _{x} h _{R/L} = i \omega \sin \alpha h _{R/L}$, the matching conditions (\ref{BC1-dw}) and (\ref{BC2-dw}) can be written as follows:
\begin{equation}
\left[ \begin{array}{c} h _{R/L} ( 0 ^{+} ) \\ h _{R/L} ^{\prime} ( 0 ^{+})  \end{array} \right] = \frac{1}{\Delta} \left[ \begin{array}{cc} 1 + ( \xi / 2) ^{2} & \mp ( \xi / \omega \cos \alpha ) \\ \mp ( \xi \omega \cos \alpha ) & 1 + ( \xi / 2 ) ^{2} \end{array} \right] \left[ \begin{array}{c} h _{R/L} ( 0 ^{-} ) \\ h _{R/L} ^{\prime} ( 0 ^{-} ) \end{array} \right] ,  \label{GeneralBC}
\end{equation}
where $\Delta = 1 - (\xi /2) ^{2}$ and $\xi \equiv {\tilde \theta} \omega ^{2} \cos ^{2} \alpha \neq 2$. As shown in the \ref{App-SAE}, this result corresponds to the choice
\begin{equation}
X _{1} = \mp  \xi \omega \cos \alpha \quad , \quad X _{4} = \pm \xi / \omega \cos \alpha \quad , \quad X _{2} = X _{3}=0 , \label{CHOICE1}
\end{equation}
which shows that the operator $\hat{\mathcal{O}} _{\rm dw}$, defined by Eq. (\ref{O-dw}), corresponds to one of the possible SAE of the operator $- d ^{2} / dz ^{2}$, according to the results in Ref.~\refcite{Kurasov}. In other words, we have chosen the boundary
conditions (\ref{GeneralBC}) for Eq. (\ref{Field-Eq-CircPol}) by demanding the operator
\begin{equation}
- \frac{d ^{2}}{dz ^{2}} \mp {\tilde \theta} \omega \cos \alpha \frac{d}{dz}\delta \left( z\right) \frac{d}{dz} \mp {\tilde \theta} \omega ^{3} \cos ^{3} \alpha \delta \left( z\right) \label{SA_FIN_OP}
\end{equation}
to be self-adjoint. We emphasize that our notation is $\frac{d}{dz} \delta \left( z \right) \frac{d}{dz} = \delta \left( z \right) \frac{d ^{2}}{dz ^{2}} + \delta ^{\prime} \left( z \right) \frac{d}{dz}$ instead of $\frac{d}{dz} \delta \left( z \right) \frac{d}{dz} = \delta ^{\prime} \left( z \right) \frac{d}{dz}$ as considered in Refs.~\refcite{Langmann} and \refcite{Gadella}.

Imposing the boundary conditions on $h _{R/L}$, we obtain
\begin{equation}
\mathcal{T} _{R/L} \left( \xi \right) = \frac{4 - \xi ^{2}}{4 + \xi ^{2}} \quad , \quad \mathcal{R} _{R/L} \left( \xi \right) = \pm i \frac{4 \xi}{4 + \xi ^{2}} ,  \label{Amplitudes}
\end{equation}
where $\xi = {\tilde \theta} \omega ^{2} \cos ^{2} \alpha \neq 2$. It can be easily verified that $\vert \mathcal{T}_{R/L} \vert ^{2} + \vert \mathcal{R}_{R/L} \vert ^{2} = 1$, which is consistent with energy conservation.

Going back to our analogy with quantum mechanics, let us recall that in one-dimensional scattering the fraction of particles that is transmitted (for an arbitrary given potential) in general vanishes at threshold, i.e. as the kinetic energy of the incident particles approaches zero. Intuitively, this occurs because the potential is acutely large as compared with the (small) energy of the incident particles. However, it has been shown that for a potential consisting of two Dirac delta functions of arbitrary strength, a finite portion of the incident particles is transmitted at threshold for certain choices of the set of parameters defining the potential \cite{Senn}. In the problem at hand, the only parameter associated with the point interaction in (\ref{O-dw}) is ${\tilde \theta}$, and clearly our result (\ref{Amplitudes}) suggests a threshold anomaly in the sense of Ref.~\refcite{Senn} for an arbitrary value of ${\tilde \theta}$. In the absence of the domain wall, our results lead to total transmission and consequently to no reflection, as expected. However this result can also be obtained in the limit $\omega \rightarrow \infty$, i.e. when the energy of the incident wave is very large as compared to the potential strength $\tilde{\theta}$. As the expressions for the reflection and transmission coefficients indicate, the effect of the domain wall becomes important when $\xi={\tilde \theta} \omega ^{2} \cos ^{2} \alpha \simeq 2$, by substantially increasing the reflecting property of the domain wall. Nevertheless, as shown in Eq. (\ref{GeneralBC}), the value $\xi=2$ is strictly forbidden, in such a way that we never have a perfect mirror  for gravitational waves.

\subsubsection{The emergence of the area matching condition}

\label{RefractiveGW}

\begin{figure}[ht]
\centerline{ \psfig{file=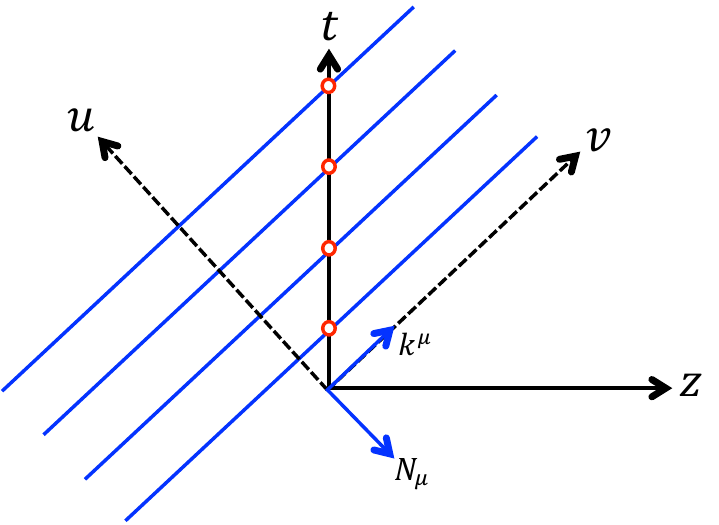,width=6cm}}
\vspace*{8pt}
\caption{{\small The case of a normal incident wave  propagating in the $+z$ direction. The parallel lines 
$u = \mbox{\small const.}$ corresponds to the phase $\Phi = z - t = \mbox{\small const.}$ The axis  coming out of the page from  the origin corresponds to  the $xy$ plane. The spacelike hypersurface $\Sigma$ is $z = 0$, with normal $n _{\mu} = (0,0,0,1)$. The  null hypersurfaces $N$ are  given by $u= \mbox{\small const.}$, with normal $N _{\mu} = (-1,0,0,1)$. The two-dimensional surfaces, where  areas are  measured at fixed time $t$, correspond to the intersection $\Sigma \cap N$ and are shown as dots on the $t$ axis}.}
\label{wave}
\end{figure}

In this section, we show that the discontinuous metric at $z=0$ which we found in the previous section,  satisfies the area matching condition on the spacelike two-dimensional surfaces arising from the intersection of  the null hypersurface $N$ describing the incident  wave and the hypersurface $\Sigma$ describing the domain wall. We illustrate the process for the case of normal incidence ($\alpha = 0$) in Fig. \ref{wave}, but we discuss the general case in the following. Let us recall that, in a given coordinate system $(\eta ^{2} , \eta ^{3})$,  the cross-sectional area $\mathcal{A}$ of any spacelike two-surface  is 
calculated as 
\begin{equation}
\mathcal{A} = \int \sqrt{\sigma} d \eta ^{2} d \eta ^{3}.  \label{Area}
\end{equation}
In our case, $\sigma$ is the determinant of the corresponding induced two-metric $\sigma _{AB}$ in $N \cap \Sigma$, and $A , B = 2 , 3$. In this way, we need to prove that $\sigma$ is continuous there even if the metric $\sigma _{AB}$ is discontinuous. As we will see in the following, the detailed expression for the discontinuous four-metric (\ref{metricGW}) is irrelevant for our purposes.

Let us recall that the null hypersurface $N$, 
determined  by the impinging gravitational wave,  is defined by the 
constant phase $\Phi={\hat k}_\mu x^\mu$, 
yielding 
\begin{equation}
\Phi = - t + x \sin \alpha + z \cos \alpha = C .  \label{NULLH}
\end{equation}
The normal vector to $N$ is $N _{\mu} = \partial \Phi / \partial x ^{\mu} = ( - 1, \sin \alpha , 0 , \cos \alpha)$. Following Ref.~\refcite{Poisson}, we introduce on the hypersurface $N$ a coordinate system $y ^{a} = ( \lambda , \eta ^{2} , \eta ^{3})$, $a=1,2,3$, in such a way that its parametrization $x ^{\mu} = (t , x , y , z) = x ^{\mu} (\lambda ,\eta ^{2} , \eta ^{3})$ satisfies
\begin{equation}
\frac{\partial x ^{\mu}}{\partial \lambda} = k ^{\mu} = e _{1} ^{\mu}\qquad , \qquad \frac{\partial x ^{\mu}}{\partial \eta ^{A}} = e _{A}^{\mu } \qquad , \qquad A=2,3,
\label{TRIAD}
\end{equation}
with $e _{A}^{\mu}$ spanning any two-dimensional subspace orthogonal to $k ^{\mu}$. The following relations 
\begin{align}
t = \lambda - C \qquad &, \qquad x = \lambda \sin \alpha + \eta ^{2} \cos \alpha , \notag \\ y = \eta ^{3} \qquad &, \qquad z = \lambda \cos \alpha - \eta ^{2} \sin \alpha,  \label{NPARAM1}
\end{align}
are a parametrization of $\Phi$ and lead to 
\begin{align}
\frac{\partial x ^{\mu }}{\partial \lambda} = ( 1 , \sin \alpha , 0 , \cos \alpha ) = k ^{\mu} , \notag \\[4pt] e _{2} ^{\mu} = ( 0 , \cos \alpha ,0 , - \sin \alpha ) \qquad , \qquad e _{3} ^{\mu} = (0,0,1,0).  \label{TRIAD1}
\end{align}
For the case of normal incidence, we have $k ^{\mu} = (1,0,0,1)$ together with $e ^{\mu} _{2} = (0,1,0,0)$ and $e ^{\mu} _{3} = (0,0,1,0)$, the later being the two basis vectors for the $xy$ plane located at  the domain wall for constant $t$. The next step is to obtain the induced two-metric $\sigma _{AB} = g _{\mu \nu} e _{A} ^{\mu} e _{B} ^{\nu}$, in $N$. Recalling the general expression 
\begin{equation}
g _{\mu \nu} = \eta _{\mu \nu} + \varepsilon \tilde{e} _{\mu \nu}^{(+)} h _{+} + \varepsilon \tilde{e} _{\mu \nu} ^{(\times)} h _{\times} ,  \label{GENMET}
\end{equation}
where the polarization tensors are given in Eq. (\ref{PolarizationTensors}), a direct calculation yields 
\begin{equation}
\left[ \sigma _{AB} \right] = \left[  \begin{array}{cc} 1 + \varepsilon h _{+} \cos ^{2} \alpha & \varepsilon h _{\times} \cos \alpha \\ \varepsilon h _{\times} \cos \alpha & 1 - \varepsilon h _{+} \cos ^{2} \alpha \end{array} \right] .  \label{2DMETRIC}
\end{equation}
Since we are considering only the linear approximation in our calculation,
we must ignore terms proportional to $\varepsilon ^{2}$. This leads to 
\begin{align}
\sigma =\det \left[ \sigma _{AB} \right] = 1 + \mathcal{O} (\varepsilon ^{2}).  \label{FINALDET}
\end{align}
That is to say, the determinant of the induced two-metric is continuous in $N$, in particular in $N \cap \Sigma$, thus erasing any discontinuous contribution to the calculation of areas of spacelike two-surfaces in $N$. 

\subsection{Gravitational waves propagating through a surface layer}
\label{CONTGW}

Here, we proceed in a similar fashion to that of the previous section. Introducing the relations (\ref{DiscFuncDef}) into the modified wave equation (\ref{Field-Eq-CircPol}) with the point interaction (\ref{O-sl}), we obtain 
\begin{align}
\Box h _{R/L}\left( z \right) &= \mp i \frac{{\tilde \theta} \cos \alpha}{d} \Big[ \Pi \left( z \right) \partial _{t} \Box h _{R/L} (z) + \delta \left( z \right) \overline{\partial _{z} \partial _{t} h _{R/L} \left( 0 \right)} - \delta \left( z - d \right) \overline{\partial _{z} \partial _{t} h _{R/L} \left( d \right)} \Big] . \label{GWs-sl}
\end{align}
Next we proceed to the calculation of the boundary conditions.

\subsubsection{Boundary conditions}

We first compute the boundary condition for the derivative of the metric perturbation at $z=0$ by integrating (\ref{GWs-sl}) over the interval $[- \varepsilon , + \varepsilon]$ and taking the limit $\varepsilon \rightarrow 0 ^{+}$, i.e.
\begin{align}
h ^{\prime} _{R/L} (0 ^{+}) & - h ^{\prime} _{R/L} (0 ^{-}) = \mp i \frac{{\tilde \theta} \cos \alpha}{d} \Big[ \lim _{\varepsilon \rightarrow 0 ^{+}} \int _{- \varepsilon} ^{+ \varepsilon} H \left( z \right) \partial _{t} \partial _{z} ^{2} h _{R/L} dz + \overline{\partial _{z} \partial _{t} h _{R/L} \left( 0 \right)} \Big] . \label{BC1.1-sl}
\end{align}
The remaining integral can be computed by parts. The result is
\begin{align}
h ^{\prime} _{R/L} (0 ^{+}) & - h ^{\prime} _{R/L} (0 ^{-}) = \mp i \frac{{\tilde \theta} \cos \alpha}{d} \partial _{z} \partial _{t} h _{R/L} (0 ^{+}) , \label{BC1-sl}
\end{align}
where we used that $H (- 0 ^{+}) = 0$. In order to obtain the boundary condition for the metric perturbation, we start by integrating Eq. (\ref{GWs-sl}) from $-L$ (with $L>0$) to $z$:
\begin{align}
h ^{\prime} _{R/L} (z) - h ^{\prime} _{R/L} (- L) - \partial _{t} ^{2} \int _{-L} ^{z} h _{R/L} (z ^{\prime}) dz ^{\prime} = \mp \frac{i {\tilde \theta} \cos \alpha}{d} \times \notag \\ \left[ \int _{-L} ^{z} H(z ^{\prime}) \Box ^{\prime} \partial _{t} h _{R/L} (z ^{\prime}) dz ^{\prime} + H(z) \overline{\partial _{z} \partial _{t} h _{R/L} \left( 0 \right)} \right] . \label{BC2.1-sl}
\end{align}
The integral in the right-hand side can be computed by parts. The result is
\begin{align}
h ^{\prime} _{R/L} (z) - h ^{\prime} _{R/L} (- L) - \partial _{t} ^{2} \int _{-L} ^{z} h _{R/L} (z ^{\prime}) dz ^{\prime} = \mp \frac{i {\tilde \theta} \cos \alpha}{d} H (z) \partial _{z} \partial _{t} h _{R/L} (z) . \label{BC2.2-sl0}
\end{align}
Next we integrate the above equation with respect to $z$ from $- \varepsilon$ to $+ \varepsilon$. Taking the limit $\varepsilon \rightarrow 0 ^{+}$, we obtain
\begin{align}
h _{R/L} (0 ^{+}) - h _{R/L} (0 ^{-}) &= \mp \frac{i {\tilde \theta} \cos \alpha}{d} \left[ \partial _{t} h _{R/L} (0 ^{+}) - \overline{ \partial _{t} h _{R/L} (0)} \right] . \label{BC2.2-sl}
\end{align}
Since the time-dependence of all wave amplitudes is $e ^{- i \omega t}$, the only solution to Eq. (\ref{BC2.2-sl}) is
\begin{align}
h _{R/L} (0 ^{+}) = h _{R/L} (0 ^{-}) . \label{BC2-sl}
\end{align}
In other words, the metric is continuous at $z = 0$.

The boundary conditions at  $z = d$ are obtained from a calculation completely analogous to that for $z=0$, which we do not reproduce here. The results are
\begin{align}
h ^{\prime} _{R/L} (d ^{+}) - h ^{\prime} _{R/L} (d ^{-}) &= \pm \frac{i {\tilde \theta} \cos \alpha}{d} \partial _{t} h ^{\prime} _{R/L} (d ^{-}) , \label{BC1-sl-d} \\ h _{R/L} (d ^{+}) &= h _{R/L} (d ^{-}) . \label{BC2-sl-d}
\end{align}

\subsubsection{The reflection and transmission coefficients}
\label{RT_SL}

As discussed in Sec. \ref{theta-models}, the metric perturbation satisfies the usual wave equation both within and outside the surface layer. Assuming that the source lies in the region $z < 0$, yielding a  gravitational wave impinging on the surface $z = 0$ from the left, the metric perturbation can be taken as
\begin{align}
h _{R/L} = \left\lbrace \begin{array}{c} e ^{i k _{\mu} x ^{\mu}} + \mathcal{R} _{R/L} e ^{i \tilde{k} _{\mu} x ^{\mu}} \\[5pt] A _{R/L} e ^{i k _{\mu} x ^{\mu}} + B _{R/L} e ^{i \tilde{k} _{\mu} x ^{\mu}} \\[5pt] \mathcal{T} _{R/L} e ^{i k _{\mu} x ^{\mu}} \end{array} \right. \begin{array}{l} ; \quad z < 0 \\[5pt] ; \quad 0<z<d \\[5pt] ; \quad z > d \end{array}  \label{metricGW-sl}
\end{align}
Imposing the previously found boundary conditions on the metric perturbation, we obtain the following reflection and transmission amplitudes
\begin{align}
\mathcal{R} _{R/L} = \pm \frac{ \gamma  \left[ 2 \pm  \gamma \right] \left( 1 - e^{ i \phi} \right)}{\gamma^2 e ^{ i\phi} - \left[ 2 \pm  \gamma \right] ^{2}} \qquad , \qquad
 \mathcal{T} _{R/L} = -\frac{4 \left[ 1 \pm  \gamma \right] }{\gamma^2 e ^{ i \phi} - \left[ 2 \pm  \gamma \right] ^{2}}, \label{RTAMP}
\end{align}
where
\begin{equation}
\gamma= \omega {\tilde \theta} \cos\alpha / d \qquad , \qquad \phi=2\omega d \cos\alpha.
\end{equation}
Let us recall that in the units we are working ${\tilde \theta}$ has dimensions of length square.
Note that the above results naturally depend on the thickness $d$ of the surface layer. One easily verifies that $\vert \mathcal{R} _{R/L} \vert ^{2} + \vert \mathcal{T} _{R/L} \vert ^{2} = 1$, which is consistent with energy conservation.

It is interesting to compute the limit when $d \rightarrow 0$. After performing a series expansion in powers of $d$ and taking the limit, we find
\begin{align}
\mathcal{R} _{R/L} = \frac{- i {\tilde \theta} \omega ^{2} \cos ^{2} \alpha }{  i {\tilde \theta} \omega ^{2} \cos ^{2} \alpha \mp 2} \qquad , \qquad \mathcal{T} _{R/L} = \frac{ \mp 2}{ i {\tilde \theta} \omega ^{2} \cos ^{2} \alpha \mp 2}.
\label{RTLINEARCASE} 
\end{align}
From the above, the transmission and reflection coefficients in this limiting case are
\begin{align}
\vert \mathcal{R} _{R/L} \vert ^{2} = \frac{ \xi ^{2}}{ 4 + \xi ^{2}} \qquad , \qquad \vert \mathcal{T} _{R/L} \vert ^{2} = \frac{4}{4 + \xi ^{2}} ,
\label{RTLINEARCASE1}
\end{align}
where $\xi = {\tilde \theta} \omega ^{2} \cos ^{2} \alpha$. 

Let us observe that the limiting results in Eqs. (\ref{RTLINEARCASE}) and (\ref{RTLINEARCASE1}) do not coincide with the corresponding values that we obtained in the case of a domain wall of $\theta$. Nevertheless, they have a similar form to those obtained in CS electrodynamics \cite{Karch, Hehl, Zanelli, Huerta} with the exception that, contrary to that case, now these coefficients depend on the frequency. This is expected since the gravitational equations of motion have additional derivatives with respect to those in electrodynamics.

\section{Some Phenomenological Estimations} \label{phenomenological}

Let us now discuss some estimations of the parameters in our model, together with their impact upon wave propagation, codified in the reflection and transmission coefficients, by restricting our general approach to some cases already considered in the literature. In this section, we go back to standard units in order to make a smooth transition to the condensed matter case.

Let us first recall the general conditions for the applicability of the linearized approximation. To being with, we are dealing with the weak field limit of Einstein equations $\vert h _{\mu \nu} \vert \ll 1$, which means that the Riemann tensor can be estimated to be $\vert R \vert \sim \omega ^{2} /c ^{2} \ll M ^{2} c ^{2} / \hbar ^{2}$, where $\omega$ is the frequency of the wave which propagates with the maximum attainable velocity $c$ in the medium and $M$ is the mass scale under which the effective theory is valid. In this approach, the CS interaction is considered as an effective theory arising from the integration of fermions in a more fundamental model valid for energies larger than the effective energy $M c ^{2}$. As a matter of fact, both Loop Quantum Gravity and String Theory unavoidable yield the CS modified gravity included in their low energy limit \cite{Alexander}. Moreover, the CS contribution is expected to be a small perturbation of the Einstein-Hilbert term, which can be codified in the ratio $[(\theta /4) \frac{1}{2}{}^{\ast} RR] / [R] \sim \omega ^{2} \theta / 8 c ^{2}$.

In order to have a unified description of the systems to be considered we
start from the action (\ref{Action}). The Einstein-Hilbert contribution is 
\begin{equation}
\mathcal{S} _{\scriptsize \mbox{EH}} = \frac{c ^{3}}{16 \pi G} \int d ^{4} x \sqrt{g} R = \frac{c ^{3}M ^{2}}{16 \pi \hbar }\int dt\;d^{3}x\sqrt{g}R ,  \label{EHUNITS}
\end{equation}
where $G = \hbar c / M^{2}$ is the effective gravitational constant. The CS action in Eq. (\ref{ACTIONS}) is parametrized as
\begin{equation}
\mathcal{S} _{\scriptsize \mbox{CS}} =\frac{c^{3}M^{2}}{64\pi \hbar }\theta \int dt\;d^{3} x \; \mathcal{P}
\label{CSMUUNITS}
\end{equation}
where the parameter $\theta$ with dimensions $\left[ \theta \right]
= \mbox{cm} ^{2}$, is to be identified in each case.

\subsection{The SECM case}

We first consider the gravitational case of Ref. \refcite{Smith} (to be called SECM for the initial of the authors), where the effect of the CS theory on bodies orbiting the earth is considered, and apply it to our surface layer model. Here, the parameters take their standard values where $M = M _{\text{Planck}}$, $G$ is the standard Newton constant and $c$ is the speed of light. Thus, in this section, it is simpler to work in units where $\hbar =c=1$. The identification of the action (\ref{CSMUUNITS}) with the corresponding one in Ref. \refcite{Smith} yields 
\begin{equation}
\theta = \frac{16 \pi}{3} \frac{l \theta _{SECM}}{M _{\text{Planck}} ^{2}} . \label{thetaSECM}
\end{equation}
The relevant parameter in Ref. \refcite{Smith} is what they call the CS mass, defined by
\begin{equation}
m _{CS} = - \frac{3M _{\text{Planck}} ^{2}}{8 \pi } \frac{1}{l \;\dot{\theta} _{SECM}} .
\label{CMMASS}
\end{equation}
This yields $\dot{\theta} = - 2 / m _{CS}$, where $\dot{\theta} = d \theta /dt$. Besides, the lower bound $m_{0} = 2 \times 10 ^{-13}$ eV, such that $m _{0} < m _{CS}$, has been established in Ref. \refcite{Smith}. This bound has been improved to $m_{0} = 4.7 \times 10 ^{-10}$ eV in Ref. \refcite{Qiang}. Now we make contact with our surface layer model where the parameter determining the size of the CS corrections to the reflection and transmission coefficients is $\gamma = \omega ( \tilde{\theta} / d ) \cos \alpha$, where we can identify $\tilde{\theta} / d$ as $d \theta / dz$ within the layer. Since Eq. (4) in Ref. \refcite{Smith} tells us that $\theta _{SECM}$ propagates at the speed of light, we can estimate $d \theta / d z = \dot{\theta} = - 2 / m _{CS}$ in such a way that 
\begin{equation}
\gamma = - 2 \frac{\omega}{m _{CS}} \; \cos \alpha .  \label{GAMMAEST}
\end{equation}
A Taylor expansion in powers of $\gamma$ produces, for example,
\begin{equation}
\vert T_{R} \vert = 1 - \frac{1}{2} (1-\cos \phi ) \gamma ^{2} + O (\gamma ^{6}) . 
\end{equation}
Taking $\gamma \sim 0.01$ generates corrections of the order of $10 ^{-4}$ in $\vert T_{R} \vert$ (for the largest value $\phi = \pi$) which yields $\omega _{\min} = \gamma m _{0} / 2 = 2. 4 \times 10 ^{-12} \, \mbox{eV} = 585.5$ Hz. Given that $\omega _{\min} / \omega _{\text{Planck}} \sim 10 ^{-40}$, we can safely increase $\omega$ to get a larger value of $\gamma$, still remaining within the weak field approximation. Also we notice that $\omega _{\min}$ is comparable with the frequencies in the interval 35-250 Hz corresponding to the recently observed gravitational waves in LIGO \cite{LIGO}.

\subsection{The condensed matter case}

The next case we consider is in the realm of TSCs and superfluids. Here, the part of the effective action which is related with the topology of the band structure of such materials corresponds to an action of the CS type, while the nontopological term is given by the standard Einstein-Hilbert action. Clearly, the effective character of those actions in the CM case must be reflected in choosing the appropriate scales corresponding to the basic parameters $G$, $c$ and $\theta$ of the action (\ref{ACTIONS}) \cite{VolovikBook}. Following Ref. \refcite{GEM-Sekine}, we take the maximum attainable velocity in the material to be the Fermi velocity $v_{F}$, in such a way that now we have the replacement $c \rightarrow v_{F}$. Also, the maximum energy scale is defined as the energy $\Delta $ of the superconductor energy gap (the difference between the ground state of the superconductor and the energy of the lowest quasiparticle excitation). This corresponds to the replacement $M \rightarrow \Delta / v _{F} ^{2}$ and yields $G \rightarrow G _{CM} = \hbar v _{F} ^{5} / \Delta ^{2}$.

One of the main objectives in this CM situation is to determine the thermal response of such materials. In very general terms, the idea is that a temperature gradient in the $z$-direction, for example, would produce a thermal current in the $x-y$ plane which effects as a mechanical rotation with angular velocity in the $z$-direction could be detected experimentally \cite{GEM-Sekine}. This is completely analogous to the magnetoelectric effect in TIs, whereby an electric field in the $z$-direction induces an electrical (Hall) current in the $x-y$ plane, which in turn generates a magnetic field in the $z$-direction. A precise way of dealing with this approach is to push further the electromagnetic analogy by introducing the gravitoelectromagnetic (GEM) approximation in the effective gravitational theory described by the actions $\mathcal{S} _{\scriptsize \mbox{EH}} + \mathcal{S} _{\scriptsize \mbox{CS}}$ with the corresponding parameters $v _{F}$ and $\Delta$. The details are given in Refs.~\refcite{TSC-Ryu} and \refcite{GEM-Furusaki}-\refcite{Luttinger}.

What is relevant for us is that gravitoelectromagnetism, being a weak field approximation already pertinent to CM physics, serves to motivate the study of the complementary sector which describes wave propagation. To this end, we will restrict our general action \ref{ACTIONS}, with the corresponding change of scales, to the proposed actions already considered in the CM literature in order to estimate the impact upon the reflected and transmitted waves in the domain wall case.

\subsubsection{The purely topological CS action $\mathcal{S} _{CM}$}

The corresponding action is given by Eq. (\ref{ActTop}), where $\theta _{CM} = \pi$ describes a nontrivial topological contribution. We identify the values of $\theta$ and $\xi$ (at normal incidence) entering in Eq. (\ref{Amplitudes}), for the domain wall model, as 
\begin{equation}
\theta = \frac{1}{12} \frac{\hbar ^{2} v _{F} ^{2}}{\Delta ^{2}} \quad , \quad  \xi = \frac{1}{12} \frac{\hbar ^{2}}{\Delta ^{2}} \omega ^{2} = \frac{1}{12} \left( \frac{\omega}{\omega _{\max}} \right) ^{2},  \label{THETACM1}
\end{equation}
where we have introduced $\omega _{\max} = \Delta / \hbar = 1.1 \times 10 ^{12}$ Hz. For a typical topological superconductor such as Cu${}_{x}$Bi${}_{2}$Se${}_{3}$, the experimental values of the required quantities can be estimated as \cite{GEM-Sekine}
\begin{equation}
\Delta = 7 \times 10 ^{-4} \mbox{eV} = 1.12 \times 10 ^{-15} \mbox{ergs} \qquad  ,  \qquad v _{F} = 5 \times 10 ^{7} \mbox{cm/s},  \label{EXPVAL}
\end{equation}
at $T = 3K$. As in the previous case we take $\xi = 0.01$ (corrections of order $10 ^{-4}$) which yields $\omega = 0.35 \; \omega _{\max}$, which is very close to the maximum frequency where the linear approximation ceases to be valid. Taking an upper limit $\omega / \omega _{\max} < 10 ^{- 2}$ and assuming the validity of the linear approximation we obtain $\xi = 8.3 \times 10 ^{- 6}$, which produces corrections of the order $10 ^{- 10}$ in the reflection and transmission coefficients.

\subsubsection{The gravitoelectromagnetic case}

A second possibility to deal with the thermal effects in TSCs is encoded in the modified CS action \cite{GEM-Nomura}
\begin{equation}
\mathcal{S} _{\scriptsize \mbox{GEM}} = \frac{k _{B} ^{2} T ^{2}}{24 \hbar v _{F}} \int dt d ^{3}x \;\theta _{\scriptsize \mbox{GEM}} (\mathbf{x},t)\;\mathbf{E}_{g}\cdot \mathbf{B}_{g} , \label{SEKINE}
\end{equation}
which is adopted in complete analogy with the electromagnetic case, where $\mathbf{E} _{g}$ and $\mathbf{B} _{g}$ are the gravitoelectric and gravitomagnetic fields, respectively. As a first approximation, we take $\theta _{\scriptsize \mbox{GEM}} (\mathbf{x},t) = \pi $. Let us remark that the above action does not have a precise topological content and that has not been already derived from a corresponding microscopic action by fermion integration \cite{GEM-Furusaki, GEM-Sekine, Stone}. Moreover, the GEM limit of the CS action in Eq. (\ref{ACTIONS}) produces two additional derivatives with respect to the analogous electromagnetic action. In Ref. \refcite{Smith}, it is shown that the weak field approximation produces, schematically,
\begin{equation}
{}^{\ast} RR \rightarrow - 16 \partial E _{g} \partial B _{g} \qquad , \qquad \left[ E _{g} \right] = \left[ B _{g} \right] = \frac{1}{\mbox{cm}}.
\end{equation}
For the sake of estimations, we are not considering the explicit form of the
limit. In order to compare our action (\ref{ACTIONS}) with (\ref{SEKINE}), we
need to assess the effect of the two additional derivatives. We do this by
taking
\begin{equation}
\partial ^{2} \sim \frac{1}{ L _{CM} ^{2}} \quad , \quad L _{CM} = \sqrt{\frac{\hbar G _{CM}}{ v _{F} ^{3}}} , 
\end{equation}
in such a way that the action (\ref{ACTIONS}) reads
\begin{equation}
\mathcal{S} _{\scriptsize \mbox{CS}} = - \frac{\Delta ^{4}}{4 \pi \hbar ^{3} v _{F} ^{3}} \theta \int dt d ^{3}x E _{g} B_{g} ,
\end{equation}
which compared with (\ref{SEKINE}) produces, taking the absolute value, the following expression for $\theta$ and $\xi$ in the domain wall model:
\begin{align}
\theta = \frac{\pi ^{2}}{6} v _{F} ^{2} \frac{k _{B} ^{2} T ^{2}}{\Delta ^{2}}\frac{1}{\omega _{\max} ^{2}} \quad , \quad \xi = \frac{\pi ^{2}}{6}\frac{k _{B} ^{2} T ^{2}}{\Delta ^{2}} \left( \frac{\omega}{\omega _{\max }}\right)^{2} .
\end{align}
Choosing $\xi = 0.01$ and using the experimental values given in (\ref{EXPVAL}), we obtain $\omega = 2.9 \omega _{\max}$ which is certainly beyond the linear approximation. Taking an upper limit $\omega /\omega _{\max} < 10 ^{- 2}$ and assuming the validity of the linear approximation, we have $\xi = 2.2 \times 10 ^{-1}$ which produces corrections of the order $10 ^{- 2}$ in the reflection and transmission coefficients. Again, the corrections are rather small, though eight orders of magnitude above the previous case.

\section{Summary and Final Comments}  \label{Discussion}

In this paper, we have studied the propagation of gravitational waves in nondynamical CS gravity, which is defined in the action (\ref{ActTop}) by the coupling of a scalar field $\theta$ to the gravitational Pontryagin invariant $ {}^{\ast} R _{\phantom{\sigma} \tau }^{\sigma \phantom{\tau} \mu \nu} R ^{\tau} _{\phantom{\tau} \sigma \mu \nu}$. We demonstrate that the resulting modified wave equation (\ref{Field-Eq-Fin}) is consistent with the choice of the TT gauge corresponding to $\partial_\mu \overline{h}^{\mu \nu}=0$, $\overline{h}^{\mu}_{{\phantom{\mu}}\mu}=0$ and $\overline{h} ^{3 \nu}=0$ for the trace-reversed metric perturbation $\overline{h} _{\mu \nu}$. Motivated by their relevance in CM physics, we have considered two nondynamical models for the CS coupling field $\theta$: (i) a planar domain wall defined by $z=0$ and (ii) a planar surface layer where $\theta$ changes linearly across a thickness $d$. While the former is employed for  the description of the surface of the ${}^{3}$He-B phase in the presence of surface magnetization, the latter can be useful to analyze the magnetic dipole response of a topological singlet superconductor.

For both models, the field equations couple the amplitudes of the two independent linear polarization modes and include distributional contributions like $\delta ^{\prime}(z)$ and/or $\delta (z)$, as shown in Eq. (\ref{Field-Eq-Gen}). This equation is decoupled by introducing the right- and left-handed circularly polarized modes $h _{R}$ and $h _{L}$, respectively. Since the Cotton-York tensor is supported only at the interfaces of the $\theta$-models, i.e. (i) at the domain wall and (ii) at the faces of the surface layer, the $\theta$-boundaries act as an effective thin shell of matter depending on  the components of the gravitational field itself. Therefore, the bulk metrics satisfy the standard Einstein equations and they should be properly joined at the interfaces in order to provide a valid solution of the modified field equations.

To determine the boundary conditions for the metric perturbation, we have adopted the rigorous distributional approach introduced in Ref. \refcite{Kurasov} to analyze generalized point interactions in nonrelativistic quantum mechanics. In \ref{App-SAE} we review the basics of 1D point interactions considered as self-adjoint extensions of the operator $- d ^{2} / dz^{2}$. Following the distributional approach, we obtain the wave equations (\ref{GWs-dw}) and (\ref{GWs-sl}) describing the propagation of gravity waves through a domain wall and a surface layer, respectively. To this end, we need to introduce the relations (\ref{DiscFuncDef}) for the product of a discontinuous function times $\delta$ and $\delta ^{\prime}$, which arise from the approach in Ref. \refcite{Kurasov}. The boundary conditions are then obtained by integrating the equations of motion (\ref{GWs-dw}) and (\ref{GWs-sl}) over a pill-shaped region across the interfaces. We also demonstrate that the additional contribution to the field equations, i.e. the Cotton-York tensor, corresponds to one of the self-adjoint extensions of the operator $- d ^{2} / dz^{2}$, allowing us to verify the previously obtained boundary conditions. 

We find many subtleties when we deal with the propagation of gravity waves through a domain wall. In this case, the linearized field equations include distributional contributions of the type $\delta (z)$ and $\delta ^{\prime} (z)$, which arise from the fact that the CS coupling field $\theta$ is piecewise constant and the gravitational Pontryagin invariant contains second-order derivatives of the metric perturbation. Our main finding in this case is that the boundary conditions imply that both the metric and its first derivative become discontinuous at the interface. Nevertheless, as shown in Sec. \ref{RefractiveGW}, the metric naturally satisfies the area matching condition introduced by Barrett \cite{Barrett}, in which the junction condition requiring the continuity of the metric at a given hypersurface is replaced by the weaker condition that the area of any 2-surface gives a unique result when measured from each side of the hypersurface. As the expressions for the  reflection and transmission coefficients indicate, the effect of the domain wall would becomes important when $\xi = \tilde{\theta} \omega ^{2} \cos ^{2} \alpha \simeq 2$, by substantially increasing the reflecting property of the domain wall. Nevertheless, as shown in Eq. (\ref{GeneralBC}), the value $\xi=2$ is strictly forbidden, in such a way that we never have a perfect mirror for gravitational waves due to the domain wall. This case also presents the following bizarre property: the gravitational waves suffer from the threshold anomaly, which means that the transmitted amplitude does not vanish when the frequency  goes to zero. This behavior is also present in some cases of potential scattering in 1D quantum mechanics \cite{Senn}. A similar phenomenon has also been measured in some nucleus-nucleus scattering processes in nuclear physics \cite{Brandan}. A deeper understanding of this feature is beyond the scope of the present paper.

In the surface layer case, the linearized field equations include distributional contributions of $\delta$-type in each face of the layer. From the distributional point of view, this problem closely resembles the nonrelativistic quantum-mechanical delta potential. In this case, the metric perturbation becomes continuous at the surfaces, with discontinuous first derivatives. This situation is far more similar to the analogous case of propagating electromagnetic waves in Maxwell-CS electrodynamics for a domain wall of $\theta$. This similarity relies in the fact that both the Cotton-York tensor (which contains second-order derivatives of the field $\theta$) in the surface layer case and the Maxwell-CS equations (which contains first-order derivatives of the field $\theta$) depend on the $\delta$ distribution, in contrast to the gravitational domain wall, for which the Cotton-York tensor contains distributional contributions of the type $\delta$ and $\delta ^{\prime}$.

As observed at the end of Sec. \ref{RT_SL}, the limiting case $d\rightarrow 0$ of the surface layer model does not reproduce the case of the domain wall, as one could naively expect because we have $\theta _{\mathrm{dw}} (z) = \lim _{d \rightarrow 0} \theta _{\mathrm{sl}} (z)$. This can be understood because the operators distinguishing both cases do not match in such limit when acting on the metric perturbation. From Eqs. (\ref{O-sl}) and (\ref{O-dw}) we have  
\begin{equation}
\hat{\mathcal{O}}_{\rm sl} h _{R/L}=\frac{i}{d}\Big\{\Pi (z)\Box + \left[ \delta (z)-\delta (z-d) \right] \partial _{z} \Big\} \partial _{t} h _{R/L}  \label{O-sl1}
\end{equation}
for the surface layer, and 
\begin{equation}
\hat{\mathcal{O}}_{\rm dw} h _{R/L} = i \left[ \delta (z)\Box + \delta ^{\prime} (z) \partial _{z} \right] \partial _{t} h _{R/L},  \label{O-dw1}
\end{equation}
for the domain wall. Recalling the distributional definitions for the product of a discontinuous function at $z=a$ times the $\delta (a)$ and $\delta ^{\prime} (a)$ distributions, Eq. (\ref{DiscFuncDef}), we conclude that
\begin{align}
\lim _{d \rightarrow 0} \hat{\mathcal{O}}_{\rm sl} h _{R/L} \neq \hat{\mathcal{O}}_{\rm dw} h _{R/L} ,
\end{align}
because from the distributional point of view
\begin{align}
\lim _{d \rightarrow 0} \frac{1}{d} \left[ \delta (z) \overline{\partial _{z} h _{R/L} (0)} - \delta (z-d) \overline{\partial _{z} h _{R/L} (d)} \right] \neq \delta ^{\prime} (z) \overline{\partial _{z} h _{R/L} (0)} - \delta (z) \overline{\partial _{z} ^{2} h _{R/L} (0)}  . 
\end{align}
We gave some estimations of the parameters in our model by restricting our general approach to some cases already considered in the literature. We first recalled the nontopological (Einstein-Hilbert) and the topological (Chern-Simons) terms of the action in standard units. They take the form of Eqs. (\ref{EHUNITS}) and (\ref{CSMUUNITS}), respectively, in terms of the maximum attainable velocity $c$ in the medium and the mass scale $M$ under which the effective theory is valid. The surface layer model was successfully compared with the results obtained in Ref. \refcite{Smith}, where the authors study the effect of the CS theory on bodies orbiting the earth. By relating the CS mass defined in Ref. \refcite{Smith} to our parameter $\gamma$, and using the lower bound for the former, we found that the minimum frequency required to obtain a deviation of the order of $10^{-4}$ in the reflection and transmission coefficients of the wave is $585$ Hz, which is comparable with the frequencies recently observed gravitational waves in LIGO. The next case we considered is in the realm of TSCs and superfluids, where we compared our results of the domain wall  model with  two different models of the CS action: (i) the full topological response theory defined by Eq. (\ref{ACTIONS}) and (ii) the GEM case defined by Eq. (\ref{SEKINE}). The latter has been systematically used in the literature on CM systems and exhibits a complete analogy with the electromagnetic case which describes the response of TIs. The effective character of those actions in the CM case must be reflected in choosing the appropriate scales corresponding to the basic parameters $G$, $c$ and $\theta$. Imposing the frequency of the propagating wave in the material to be at least $10^{-2}$ times, the maximum attainable frequency, we find corrections in the reflection and transmission coefficients of the order $10^{-10}$ for the case (i) and $10^{-2}$ for the case (ii). Our results show that the CS corrections in wave propagation become sizable when the frequency $\omega$ of the  wave satisfy the following requirements: (a) $\omega$ is close enough  to the maximum allowed frequency $\omega_{\max}$ in each particular case and  (b) $\omega$ satisfies $\omega/\omega_{\max} < 1$ so that the linearized approximation is valid. Let us observe that in the CM system so far considered we have $\omega _{\max }=1.1\times 10^{12}$ Hz, while the corresponding value in standard  gravity  is $\omega _{\max}=2.4\times 10^{42}$ Hz, the Planck frequency,  about thirty orders of magnitude higher. In this way, topological CM systems could afford a realistic possibility to experimentally probe the effects of the CS coupling.

To close, we discuss the relevance of the CS term with respect to additional combinations of the Riemann tensor that can be added to the action and  which can be at least as important as the CS term in the weak field limit, such as the quadratic term in $f(R)$-gravity \cite{FRgravity1, FRgravity2} for example. On one hand, in the standard gravitational case the CS term is singled out  as been an unavoidable contribution arising in the  low energy limit  of String Theory and Loop Quantum Gravity \cite{Alexander}. On the other hand,  in order to describe the gravitational response of topological matter, the effective field theory should be topological in nature, thus strongly restricting the possible theories which could be taken into consideration. Therefore, besides the CS gravity defined through the Pontryagin density, other options are to consider the Euler invariant and the torsion dependent Nieh-Yan density. The latter has been recently proposed as an alternative to describe the gravitational response theory of TSCs and superfluids \cite{NiehYan}. The analysis of such possibilities in the realm of wave propagation would constitute an interesting extension of the present work.

\appendix

\section{One-Dimensional Point Interactions as Self-Adjoint Extensions of
the Operator $-d^{2}/dz^{2}$}

\label{App-SAE}

Point interactions in one dimension appear frequently as a method of simplifying the description of various physical situations, making emphasis on the significant features of the problem but leaving aside a detailed description of the interaction, which can be later included to produce a more realistic solution. Point interactions, also known as Fermi pseudopotentials, are associated with a given differential equation and correspond to potentials $V_{FP}(z)$ that are nonzero only in some specific points in the line, where they become singular. The differential equation we consider here is
\begin{equation}
\left[ -\frac{d^{2}}{dz^{2}}+V_{FP}(z) \right] \psi (z)=E\psi (z),
\label{EQ1}
\end{equation}
which simplest physical version corresponds to the classical example of the $\delta $-function pseudopotential. On a formal level we can associate to this system the one-dimensional Schr\"{o}dinger operator
\begin{equation}
H = - \frac{\hbar ^{2}}{2m} \frac{d ^{2}}{dz ^{2}} + \lambda \delta \left( z\right) , \label{SchrDelta}
\end{equation}
where $\lambda $ is a real coupling constant and $z \in \mathbb{R}$. The resulting eigenvalue equation, $H\psi =E\psi $, acquires a precise meaning when converting it into a boundary value problem. Heuristically, this equation can be interpreted as consisting of the free Schr\"{o}dinger equation $- \frac{\hbar ^{2}}{2m} \psi ^{\prime \prime} = E \psi $ for $z \in \mathbb{R} \setminus \left\{ 0 \right\} $, together with the boundary conditions $\psi \left( 0 ^{+}\right) = \psi \left( 0 ^{-}\right) $ and $\psi ^{\prime} \left( 0 ^{+} \right) - \psi ^{\prime} \left( 0 ^{-}\right) = \frac{2m \lambda}{\hbar ^{2}} \psi \left( 0 \right)$ at $z=0$. The argument presented above can be phrased in rigorous terms by using the theory of distributions. In doing so, we should consider that observables in quantum mechanics are required to be self-adjoint operators. In our case, the observable is the energy, which is formally represented by the operator $H _{0} = - \frac{\hbar ^{2}}{2m}\frac{d ^{2}}{dz ^{2}}$. Each function $\psi $ in the domain of $H _{0}$ must live in the Hilbert space $\mathcal{H} = L ^{2} \left( \mathbb{R} \right)$ of functions square-integrable on $\mathbb{R}$ and be such that $\psi ^{\prime}$ is absolutely continuous at all points $z \neq 0$, satisfying also $\psi ^{\prime \prime} \in \mathcal{H}$.

We require the following two conditions in order to declare that the operator $H _{0}$ is self-adjoint. (i) To begin with, $H _{0}$ must be hermitian (or symmetric in the mathematical language), which means that we must impose the following additional boundary conditions at $z=0$, for all $\psi $ and $\varphi $ in the domain of $H _{0}$,
\begin{equation}
\int _{\mathbb{R} \setminus \{0\}} \left[ \left( H _{0} ^{\dagger} \psi \left( z \right) \right) ^{\ast} \varphi \left( z \right) - \psi ^{\ast} \left(
z \right) H _{0} \varphi \left( z \right) \right] dz = 0 , \label{SelfAdjoint}
\end{equation}
with $H _{0} ^{\dagger} = H _{0}$. By partial integration of this relation translates into the boundary condition
\begin{equation}
\left[ \psi ^{\ast} \varphi ^{\prime} - \psi ^{\ast \prime} \varphi \right] _{z = 0 ^{+}} = \left[ \psi ^{\ast} \varphi ^{\prime} - \psi ^{\ast \prime} \varphi \right] _{z = 0 ^{-}}.  \label{SelfAdjointBC}
\end{equation}
Note that this result does not require either the functions or their first derivatives to be continuous at $z=0$. Neither it requires that the boundary conditions for the function $\psi $ on the right are exactly the same as the boundary condition for the function $\varphi $ on the left in Eq. (\ref{SelfAdjointBC}). In fact, the functions $\psi (z)$, defined in $\mathbb{R} \setminus \{0\}$, which are in the domain of the adjoint $H _{0} ^{\dagger}$ are required to be continuous with continuous bounded first derivatives except at the origin, where they can have arbitrary finite discontinuities both in $\psi $ and $\psi ^{\prime}$. This last property means that the corresponding limits at $0 ^{+}\ $and at $0 ^{-}$ are finite and well defined.

In general, boundary conditions involving a function $\psi $ and its derivative $\psi ^{\prime}$ are of the form
\begin{equation}
\left[ \begin{array}{c} \psi ( 0 ^{+} ) \\ \psi ^{\prime} ( 0 ^{+} ) \end{array} \right] = \left[
\begin{array}{cc}
u_{11} & u_{12} \\
u_{21} & u_{22}%
\end{array}%
\right] \left[ \begin{array}{c} \psi ( 0 ^{-} ) \\ \psi ^{\prime} ( 0 ^{-} ) \end{array} \right] = U \left[ \begin{array}{c} \psi ( 0 ^{-} ) \\ \psi ^{\prime} ( 0 ^{-} ) \end{array} \right] ,  \label{GeneralBCSelfAdjoint}
\end{equation}%
parametrized by the complex $2 \times 2$ matrix $U \equiv \left[ u_{ij}\right]$. (ii) The second condition for an operator to be self-adjoint is that both the domain and the action of the operator acting on the right are equal to the domain and the action of the adjoint operator acting on the left. In this case, we have only one matrix $U$ for both type of functions $\psi$ and $\varphi$, and the condition (\ref{SelfAdjointBC}) translates into \cite{Langmann, Gadella}
\begin{equation}
U ^{\dagger} J U = J \qquad , \qquad J = \left[
\begin{array}{cc}
0 & 1 \\
-1 & 0%
\end{array}%
\right] .  \label{JCOND}
\end{equation}
The above equation implies that $\det (U)$ is a phase and also reduces the parametrization of $U$ from eight to four real independent numbers. The boundary conditions for the $\delta$-function pseudopotential corresponds to the simple choice $u _{11} = u _{22} = 1$, $u _{12} = 0$ and $u _{21} = 2m \lambda / \hbar ^{2}$. Thus we now interpret the $\delta$-interaction as a SAE of the operator $H _{0}$.

To summarize, the basic problem posed in general by 1D point interactions is to consistently determine the boundary conditions  which the solutions $\psi (z)$ of the associated differential equation must satisfy around the points in the line where the pseudopotential diverges. In most of the cases discussed in the literature, this point has raised many controversies which are far from been settled down \cite{Griffiths}. As mentioned above, in quantum mechanics, the operator in the left-hand side of Eq. (\ref{EQ1}) is usually the Hamiltonian of the system. This suggests a natural way to define the corresponding boundary conditions by demanding the operator to be self-adjoint. Although we are not dealing with a quantum mechanical problem, in this work we adopt the same approach to define our boundary conditions.

General SAE of the operator $-d^{2}/dz^{2}$ have been considered in Refs.~\refcite{Kurasov} and \refcite{Albeverio}, among others, leading to a generalized point interactions which depends on both, the $\delta$- and $\delta ^{\prime}$-interactions. This method of defining the required boundary conditions has the advantage of producing results which are independent of specific models of the delta distribution and its derivatives.

It is  known that the most general SAE of the operator $- d ^{2}/dz ^{2}$ is parametrized by four real independent parameters which we denote by $X _{i}$, $i=1,2,3$ and $4$. \cite{Kurasov} The application of the distributional method in Ref.~\refcite{Kurasov} to the problem of the SAE of $H _{0}$ leads to the following result for the matrix $U$:
\begin{equation}
U = e ^{-i \arg (D)} \left[
\begin{array}{cc}
\frac{(2+X_{2})^{2}-X_{1}X_{4}+X_{3}^{2}}{|D|} & -\frac{4X_{4}}{|D|} \\
\frac{4X_{1}}{|D|} & \frac{(2-X_{2})^{2}-X_{1}X_{4}+X_{3}^{2}}{|D|}%
\end{array}%
\right] ,  \label{FIN_REL_FIN}
\end{equation}
where
\begin{equation}
D = 4 + X _{1} X _{4} - X _{2} ^{2} - X _{3} ^{2} - 4 i X _{3}  \label{DEN}
\end{equation}
is in general a complex number. The above equation (\ref{FIN_REL_FIN}) corresponds to Eq. (19) in Ref.~\refcite{Kurasov} and shows the general form of the matrix $U$, being a phase times a matrix with determinant one. The explicit form of the boundary conditions is
\begin{align}
& - X _{1} \psi (0 ^{+}) + \left[ 2 + \left( X _{2} - i X _{3} \right) \right] \psi ^{\prime} (0^{+}) = X _{1} \psi (0 ^{-}) + \left[ 2 - \left( X _{2} -i X _{3} \right) \right] \psi ^{\prime} (0 ^{-}) ,  \label{BC1} \\[8pt] & \left[ 2 - \left( X _{2} + i X _{3} \right) \right] \psi (0  ^{+}) + X _{4} \psi ^{\prime} (0 ^{+}) = \left[ 2 + \left( X _{2} + i X _{3} \right) \right] \psi (0 ^{-}) - X _{4} \psi ^{\prime} (0 ^{-}) . \label{BC2}
\end{align}
Our problem now is to obtain the matrix $U$ in Eq. (\ref{FIN_REL_FIN})
starting from the formulation of the problem in terms of pseudopotentials, in order to identify the operator that arises via the Cotton-York tensor modifications to the wave equation as a SAE of $H_{0}$, which will lead to the appropriate boundary conditions for the solution. Based upon the results of Kurasov \cite{Kurasov}, we propose the following interpretation of his distributional operator, in terms of Fermi pseudopotentials defining a second-order differential equation of the standard type for the functions $\psi (z)$ defined above
\begin{align}
H \psi \equiv - \frac{d ^{2}}{dz ^{2}} \psi - X _{4} \frac{d}{dz} \delta (z) \frac{d}{dz} \psi + i X _{3} \left( \delta ^{\prime} (z) + 2 \delta (z)\frac{d}{dz} \right) \psi + X _{1} \delta (z) \psi + X _{2} \delta ^{\prime } (z) \psi = E \psi
\label{PSEUD_POT}
\end{align}
We emphasize that our notation is
\begin{equation}
\frac{d}{dz} \delta \left( z \right) \frac{d}{dz} = \delta \left( z \right) \frac{d ^{2}}{dz ^{2}} + \delta ^{\prime} \left( z \right) \frac{d}{dz} , \label{NOT}
\end{equation}
instead of $\frac{d}{dz} \delta \left( z \right) \frac{d}{dz} = \delta ^{\prime} \left( z \right) \frac{d}{dz}$ as considered in Ref.~\refcite{Langmann}. Next, we need to make sense of products like $\psi (z)\delta (z)$ and $\psi(z) \delta ^{\prime} (z)$. For continuous functions $\varphi (z)$, with continuous first derivatives, at $z=0$, the following properties are well-known \cite{Langmann}
\begin{align}
\label{deltaf11}
\begin{split}
\varphi \left( z \right) \delta \left( z \right) &= \varphi \left( 0 \right) \delta \left( z \right) , \\[4pt] \varphi \left( z \right) \delta ^{\prime}\left( z \right) &= \varphi \left( 0 \right) \delta ^{\prime} \left( z \right) - \varphi ^{\prime} \left( 0 \right) \delta \left( z \right) . 
\end{split}
\end{align}
The generalization of the above equations to the case of the functions $\psi (z)$ defined in $\mathbb{R} \setminus \{0\}$ and  arising from the distributional construction of Ref.~\refcite{Kurasov}, consists in replacing the value of $\varphi \left(0\right) $ (and of $\varphi ^{\prime }\left( 0\right) $) by their mean value at the origin
\begin{equation}
\overline{\psi \left( 0 \right)} = \frac{1}{2} \left[ \psi \left( 0 ^{+}\right) + \psi \left( 0 ^{-} \right) \right] , \label{AverageOrigin1}
\end{equation}
where $\psi \left( 0 ^{+} \right)$ and $\psi \left( 0 ^{-} \right)$ are respectively the limits of $\psi \left( z \right) $ when $z$ approaches $0$ from the positive and negative sides. Then, the definitions (\ref{deltaf11}) should be read as
\begin{align}
\label{deltaf22}
\begin{split}
\psi \left( z \right) \delta \left( z \right) &= \overline{\psi \left( 0 \right)} \delta \left( z \right) , \\[4pt] \psi \left( z \right) \delta ^{\prime} \left( z \right) &= \overline{\psi \left( 0 \right)} \delta ^{\prime} \left( z \right) - \overline{\psi ^{\prime} \left( 0 \right)}\delta \left( z \right) .
\end{split}
\end{align}
Inserting the expressions (\ref{deltaf22}) in (\ref{PSEUD_POT}) we obtain
\begin{align}
H \psi &= - \frac{d ^{2} \psi}{dz ^{2}} + \left[ X _{1} \overline{\psi \left( 0 \right)} - \left( X _{2} - i X _{3} \right) \overline{\psi ^{\prime}\left( 0 \right)} \right] \delta (z) \nonumber \\ & +\left[ \left( X _{2}+ i X _{3} \right) \overline{\psi \left( 0 \right)} - X _{4} \overline{\psi ^{\prime} \left( 0 \right)} \right] \delta ^{\prime}(z) =
E \psi . \label{EQ_FIN}
\end{align}
Note that the dependence on $\psi ^{\prime \prime} \left( 0 ^{+}\right)$ and $\psi ^{\prime \prime} \left( 0 ^{-} \right)$ has canceled between the contributions arising from (\ref{NOT}) and the substitution (\ref{deltaf22}) when $\psi \left( z \right) \rightarrow \psi ^{\prime} \left( z \right)$. Following Ref.~\refcite{Albeverio}, we recover the boundary conditions determined by $U$ in Eq. (\ref{FIN_REL_FIN}) dealing with Eq. (\ref{EQ_FIN}) in the standard way. First, we integrate the equation from $- \varepsilon$ to $+ \varepsilon$ using
\begin{align}
\int _{- \varepsilon} ^{+ \varepsilon} dz \delta (z) = 1 \qquad , \qquad \int _{- \varepsilon} ^{+ \varepsilon} dz \delta ^{\prime} (z) =0 .   \label{BASIC_INT}
\end{align}
The result is
\begin{align}
\left[ \psi ^{\prime} (0 ^{+}) - \psi ^{\prime } (0 ^{-}) \right] - X _{1}\overline{\psi \left( 0 \right)} + \left( X _{2} - i X _{3} \right) \overline{\psi ^{\prime} \left( 0 \right)} = 0
\end{align}
which we can explicitly rewrite as
\begin{align}
& \left[ - 2 + i X _{3} - X _{2} \right] \psi ^{\prime} (0 ^{+}) + X _{1} \psi (0 ^{+}) + \left[ 2 + i X _{3} - X _{2} \right] \psi ^{\prime} (0 ^{-}) + X _{1} \psi (0 ^{-}) = 0. \label{INT1}
\end{align}
The above equation  is precisely the boundary condition in Eq. (\ref{BC1}). The remaining boundary condition is obtained integrating Eq. (\ref{EQ_FIN}) from $- L$ to positive $z$, and further from $- \varepsilon $ to $+ \varepsilon$. In this way, we have
\begin{align}
E \int _{- L} ^{z} \psi(z ^{\prime}) dz ^{\prime} &= - \left[ \psi ^{\prime } (z) - \psi ^{\prime} (L) \right] + \left[ \left( X _{2} + i X _{3} \right) \overline{\psi \left( 0 \right)} - X _{4} \overline{\psi ^{\prime} \left( 0 \right)} \right] \delta (z) \notag \\ & + \left[ X _{1} \overline{\psi \left( 0 \right)} - \left( X _{2} - i X _{3} \right) \overline{\psi ^{\prime} \left( 0 \right)} \right] H (z) , \label{INTLX}
\end{align}
where $H (z)$ is the Heaviside function. Performing the second integration in $z$ from $- \varepsilon$ to $+ \varepsilon$, we obtain
\begin{align}
& \left[ 2 - \left( X _{2} + i X _{3} \right) \right] \psi (0 ^{+}) + X _{4} \psi ^{\prime} (0 ^{+}) = \left[ 2 + \left( X _{2} + i X _{3} \right) \right] \psi (0 ^{-}) - X _{4} \psi ^{\prime} (0 ^{-}) ,  \label{INT2}
\end{align}
in the limit, since the discontinuities produced by $	H (z)$ are finite.
Equation (\ref{INT2}) reproduces the boundary conditions in Eq. (\ref{BC2}). To summarize, we have regained the conditions for the SAE of the operator $- d ^{2} / dz ^{2}$ by providing an interpretation in terms of pseudopotentials (given by Eq. (\ref{PSEUD_POT})) of the distributional operator in Ref.~\refcite{Kurasov}, plus the use of the relations (\ref{deltaf22}).

Next we apply these general results to our problem. Clearly, there is a close analogy between the quantum-mechanical case and the linearized
CS gravity we have presented in Sec. \ref{GW}. In the problem at hand, defined by Eq. (\ref{Field-Eq-CircPol}), the basic hermitian operator is the same as in quantum mechanics, $H _{0} = - \frac{d ^{2}}{dz ^{2}}$, with $E = \omega ^{2} \cos ^{2} \alpha$.
From Eq. (\ref{Field-Eq-CircPol}), we read the CS modified wave equation
\begin{equation}
\left[ -\frac{d^{2}}{dz^{2}}\mp {\tilde \theta} \omega \cos \alpha \frac{d}{dz}\delta \left( z\right) \frac{d}{dz}\mp {\tilde \theta} \omega ^{3} \cos ^{3} \alpha \; \delta \left( z \right) \right] h _{R/L} = \omega ^{2} \cos ^{2} \alpha \; h _{R/L}.  \label{CSMWE}
\end{equation}
Now we show that our CS point interaction is a particular case of the general SAE of the operator $H _{0}$. In fact, comparing Eq. (\ref{CSMWE}) with Eq. (\ref{PSEUD_POT}), we identify
\begin{equation}
X _{1} = \mp {\tilde \theta} \omega ^{3} \cos ^{3} \alpha \;\; , \;\;  X _{4} = \pm {\tilde \theta} \omega \cos \alpha \;\; , \;\; X _{2} = X _{3} = 0 ,  \label{PARAM}
\end{equation}
in such a way that the matrix U determining the boundary conditions reads
\begin{equation}
U = \frac{1}{1- \left( \xi / 2 \right) ^{2}} \left[ \begin{array}{cc} 1 + \left( \xi / 2 \right) ^{2} & \quad  \mp \left( \xi / (\omega \cos \alpha) \right) \\ \mp \left( \xi \omega \cos \alpha \right) & 1 + \left( \xi / 2 \right) ^{2} \end{array} \right] ,  \label{FINALU}
\end{equation}%
where $\xi = {\tilde \theta} \omega ^{2} \cos ^{2} \alpha \neq 2$. One can further verify that $U ^{\dagger} J U = J$. Therefore, we have demonstrated that, to linear order, the CS contribution to the wave equation for a domain wall of $\theta$, can be described as a SAE of the 1D operator $-d^{2}/dz^{2}$, with boundary conditions determined by the matrix $U$ in Eq. (\ref{FINALU}), according to the relations (\ref{GeneralBCSelfAdjoint}).

\section*{Acknowledgements}

Useful discussions with M. Cambiaso, L. Huerta, A. Toloza and J. Zanelli at an early stage of this work are warmly appreciated. Thanks are also due to C. Chryssomalakos for providing valuable references and to E. Nahmad for updating our knowledge in differential geometry. This work was supported in part by the project No. IN104815 from Direcci\'{o}n General Asuntos del Personal Acad\'{e}mico (Universidad Nacional Aut\'{o}noma
de M\'{e}xico) and CONACyT (M\'{e}xico) No. 237503.


\begin{thebibliography}{00}

\bibitem{Jackiw} R. Jackiw and S.-Y. Pi, \textit{Phys. Rev. D} \textbf{68} (2003) 104012.

\bibitem{Smith} T. L. Smith, A. L. Erickcek, R. R. Caldwell and M. Kamionkowski, \textit{Phys. Rev. D} \textbf{77} (2008) 024015.

\bibitem{Lue} A. Lue, L. M. Wang and M. Kamionkowski, \textit{Phys. Rev. Lett.} \textbf{83} (1999) 1506.

\bibitem{Yunes} S. Alexander and N. Yunes, \textit{Phys. Rev. D} \textbf{75} (2007) 124022.

\bibitem{Qi-Rev} X.-L. Qi and  S.-C. Zhang, \textit{Rev. Mod. Phys.} \textbf{83} (2011) 1057.

\bibitem{Hasan-Rev} M. Z. Hasan and C. L. Kane, \textit{Rev. Mod. Phys.} \textbf{82} (2010) 3045.

\bibitem{Peccei} R. D. Peccei and H. R. Quinn, \textit{Phys. Rev. Lett.} \textbf{38} (1977) 1440.

\bibitem{Wilczek} F. Wilczek, \textit{Phys. Rev. Lett.} \textbf{58} (1987) 1799.

\bibitem{Qi} X.-L. Qi, R. Li, J. Zang and S.-C. Zhang, \textit{Science} \textbf{323} (2009) 1184.

\bibitem{Karch} A. Karch, \textit{Phys. Rev. Lett.} \textbf{103} (2009) 171601.

\bibitem{Hehl} Y. N. Obukhov and F. W. Hehl, \textit{Phys. Lett.} \textbf{A341} (2005) 357.

\bibitem{Zanelli} L. Huerta and J. Zanelli, \textit{Phys. Rev. D} \textbf{85} (2012) 085024.

\bibitem{Huerta} L. Huerta, \textit{Phys. Rev. D} \textbf{90} (2014) 105026.

\bibitem{MCU1} A. Mart\'{i}n-Ruiz, M. Cambiaso and L. F. Urrutia, \textit{Phys. Rev. D} \textbf{92} (2015) 125015.

\bibitem{MCU2} A. Mart\'{i}n-Ruiz, M. Cambiaso and L. F. Urrutia, \textit{Phys. Rev. D} \textbf{93} (2016) 045022.

\bibitem{MCU3} A. Mart\'{i}n-Ruiz, M. Cambiaso and L. F. Urrutia, \textit{Europhys. Lett.} \textbf{113} (2016) 60005.

\bibitem{MCU4} A. Mart\'{i}n-Ruiz, M. Cambiaso and L. F. Urrutia, \textit{Phys. Rev. D} \textbf{94} (2016) 085019.

\bibitem{TSC-Wang} Z. Wang, X.-L. Qi, and S.-C. Zhang, \textit{Phys. Rev. B} \textbf{84} (2011) 014527.

\bibitem{TSC-Ryu} S. Ryu, J. E. Moore and A. W. W. Ludwig, \textit{Phys. Rev. B} \textbf{85} (2012) 045104.

\bibitem{TSC-Qi} X.-L. Qi, E. Witten and S.-C. Zhang, Phys. Rev. B \textbf{87} (2013) 134519.

\bibitem{GEM-Furusaki} A. Furusaki, N. Nagaosa, K. Nomura, S. Ryu and T. Takayanagi, \textit{C. R. Physique} \textbf{14} (2013) 871.

\bibitem{GEM-Nomura} K. Nomura, S. Ryu, A. Furusaki and N. Nagaosa, \textit{Phys. Rev. Lett.} \textbf{108} (2012) 026802.

\bibitem{GEM-Shiozaki} K. Shiozaki and S. Fujimoto, \textit{Phys. Rev. B} \textbf{89} (2014) 054506.

\bibitem{GEM-Sekine} A. Sekine, \textit{Phys. Rev. B} \textbf{93} (2016) 094510.

\bibitem{Luttinger} J. M. Luttinger, \textit{Phys. Rev.} \textbf{135} (1964) A1505.

\bibitem{Israel} W. Israel, \textit{Phys. Rev. D} \textbf{15} (1977) 935. 

\bibitem{Taub} A. H. Taub, \textit{J. Math. Phys.} \textbf{21} (1980) 1423.

\bibitem{Raju} C. K. Raju, \textit{J. Phys. A: Math. Gen.} \textbf{15} (1982) 1785.

\bibitem{Geroch} R. Geroch and J. Traschen, \textit{Phys. Rev. D} \textbf{36} (1987) 1017.

\bibitem{Clarke} C. J. S. Clarke and T. Dray, \textit{Class. Quant. Grav.} \textbf{4} (1987) 265.

\bibitem{Hogan} P. A. Hogan, \textit{Phys. Rev. Lett.} \textbf{70} (1993) 117.

\bibitem{Letelier} P. S. Letelier and A. Wang, \textit{J. Math. Phys.} \textbf{36} (1995) 3023.

\bibitem{Podolsky} J. Podolsk\'{y} and K. Vesel\'{y}, \textit{Phys. Lett. A} \textbf{241} (1998) 145.

\bibitem{Mars} M. Mars and J. M. M. Senovilla, \textit{Class. Quantum Grav.} \textbf{10} (1993) 1865.

\bibitem{Kurasov} P. Kurasov, \textit{J. Math. Anal. Appl.} \textbf{201} (1996) 297.

\bibitem{Albeverio} S. Albeverio, L. Dabrowsi and P. Kurasov, \textit{Lett. Math. Phys.} \textbf{45} (1998) 33.

\bibitem{Barrett} J. W. Barrett, \textit{Refractive gravitational waves and
quantum fluctuations}, arXiv: gr-qc/001105.

\bibitem{VolovikBook}  G. E. Volovik, \textit{The Universe in a Helium Droplet}, (Clarendon, Oxford, 2003).

\bibitem{Schutz} B. F. Schutz, \textit{A First Course in General Relativity}, 2nd ed. (Cambridge University Press, Cambridge, 2009).

\bibitem{Padmanabhan} T. Padmanabhan, \textit{Gravitation: Foundations and Frontiers}, 1st ed. (Cambridge University Press, Cambridge, 2010).

\bibitem{Griffiths} D. J. Griffiths, \textit{J. Phys. A: Math. Gen.} \textbf{26} (1993) 2265.

\bibitem{Coutinho} F. A. B. Coutinho, Y. Nogami and J. F. Perez, \textit{J. Phys. A: Math. Gen.} \textbf{30} (1997) 3937.

\bibitem{Seba} P. \v{S}eba, \textit{Rep. Math. Phys.} \textbf{24} (1986) 111.

\bibitem{Christiansen} P. L. Christiansen, H. C. Arnbak, A. V. Zolotaryuk, V. N. Ermakov and Y. B. Gaididei, \textit{J. Phys. A: math Gen.} \textbf{36} (2003) 7589.

\bibitem{Toyama} F. M. Toyama and Y. Nogami, \textit{J. Phys. A: Math. Theor.} \textbf{40} (2007) F685.

\bibitem{Langmann} M. Halln\"{a}s, E. Langmann and C. Paufler, \textit{J. Phys. A: Math. Gen.} \textbf{38} (2005) 4957.

\bibitem{Gadella} M. Gadella, J. Negro and L. M. Nieto, \textit{Phys. Lett. A} \textbf{373} (2009) 1310.

\bibitem{Senn} P. Senn, \textit{Am. J. Phys.} \textbf{56} (1988) 916.

\bibitem{Poisson} E. Poisson, \textit{A Relativist's Toolkit: The Mathemathics of Black-Hole Mechanics}, (Cambridge University Press, Cambridge, 2004).

\bibitem{Brandan} M. E. Brandan and G. E. Satchler, \textit{Phys. Rep.} \textbf{285} (1997) 143.

\bibitem{Alexander} S. Alexander and N. Yunes, \textit{Phys. Rev. D} \textbf{77} (2008) 124040.

\bibitem{Qiang} Li. Qiang and P. Xu, \textit{Gen. Rel. Grav.} \textbf{47} (2015) 26.

\bibitem{LIGO} B. P. Abbott et al., \textit{Phys. Rev. Lett.} \textbf{116} (2016) 061102.

\bibitem{Stone} M. Stone, \textit{Phys. Rev. B} \textbf{85} (2012) 184503.

\bibitem{FRgravity1} T. P. Sotiriou and V. Faraoni, \textit{Rev. Mod. Phys.} \textbf{82} (2010) 451.

\bibitem{FRgravity2} A. De Felice and S. Tsujikawa, \textit{Living Rev. Relativ.} \textbf{13} (2010) 3.

\bibitem{NiehYan} Y. Hidaka, Y. Hirono, T. Kimura and Y. Minami, \textit{Prog. Theor. Exp. Phys.} \textbf{2013} (2013) 013A02.


\end{thebibliography}
\end{document}